\documentstyle[11pt]{article}

\newcommand{\be}{\begin{equation}}
\newcommand{\ee}{\end{equation}}
\newcommand{\bea}{\begin{eqnarray}}
\newcommand{\eea}{\end{eqnarray}}
\newcommand{\ear}{\end{array}}
\newcommand{\non}{\nonumber}
\newcommand{\ra}{\rangle}
\newcommand{\la}{\langle}

\newcommand{\al}{\alpha} 
\newcommand{\bt}{\beta} 
\newcommand{\gm}{\gamma} 
\newcommand{\dt}{\delta} 

\begin{document}


\title{Introduction to solvable lattice models 
in statistical and mathematical physics 
\footnote{
In "Classical and Quantum Integrable Systems: Theory and
Applications" (ISBN 07503 09598), Institute of Physics Publishing, (2003)
http://bookmarkphysics.iop.org/ (Chap. 5 , pp. 113-151.)
%
}}  

\author{Tetsuo Deguchi
\footnote{e-mail deguchi@phys.ocha.ac.jp}}

\maketitle

\begin{center}
 Department of Physics, Ochanomizu University, \\
2-1-1 Ohtsuka, Bunkyo-ku,Tokyo 112-8610, Japan
\end{center} 
\date{}

\begin{abstract}
Some features of integrable lattice models 
are reviewed for the case of the six-vertex model. 
By the Bethe ansatz method we derive the free energy 
of the six-vertex model. 
Then, from the expression of the free energy 
we show analytically  the critical singularity near the  
phase transition in the anti-ferroelectric regime, 
where the essential singularity similar to the Kosterlitz-Thouless   
transition appears. 
We discuss the connection of the six-vertex model 
to the conformal field theory with $c=1$. 
We also introduce 
various exactly solvable models defined on two-dimensional lattices 
such as the chiral Potts model and the IRF models.  
We show that the six-vertex model has 
rich mathematical structures such as the quantum groups 
and the braid group.   
\par 
The graphical approach is emphasized in this review. We 
explain the meaning of the Yang-Baxter equation by its diagram.  
Furthermore, we can understand the defining relation  of the 
algebraic Bethe ansatz by the graphical representation. 
We can thus easily translate 
formulas of the algebraic Bethe ansatz  
into those of the statistical models. 
As an illustration, we show explicitly how we can derive  
Baxter's expressions from those of the algebraic Bethe ansatz.    
\end{abstract} 

\par \noindent 
{\bf Keywords:} exactly solvable models, the Yang-Baxter equation, statistical mechanics, 
phase transitions, the Bethe ansatz,  the six-vertex model, 
integrable lattice models, conformal field theory, two-dimensional lattice


%
%

\setcounter{equation}{0} 
\setcounter{figure}{0}
\renewcommand{\theequation}{1.\arabic{equation}}
\renewcommand{\thefigure}{1.\arabic{figure}}
\setcounter{section}{0}

\section{Introduction}

\par 
We introduce the six-vertex model defined 
on a two-dimensional square lattice. 
We describe the model in detail,  
since it gives an important  prototype of many  solvable lattice models 
defined on two-dimensional lattices \cite{Baxter-book}. 
The transfer matrix of the six-vertex model 
generalizes the XXZ quantum spin chain which plays a central role 
among integrable quantum spin chains \cite{Bethe,Yang-Yang}.     
The eight-vertex model, 
which generalizes the six-vertex model directly,   
may be considered as the most important exactly solvable model
in statistical mechanics \cite{Baxter72}.  
Moreover, many mathematical theories such 
as the algebraic Bethe ansatz \cite{8VABA} and 
 quantum groups \cite{Drinfeld,Jimbo}
are closely related to the six-vertex model.  
Starting from the six-vertex model, 
one may have a wide viewpoint on  
various physical and mathematical 
topics related to solvable models.  
There are quite a large number of topics related to exactly solvable models 
in physics and mathematics 
\cite{Baxter-book,McCoy-Wu,Ziman,Gaudin,Reprints-CFT,Itzykson,braid,Jimbo-reprints,Martin,Korepin,Lusztig-book,Chari-Pressley,Reprints-Hubbard,Jimbo-Miwa,Tsvelik-book,Majid,DMS,Etingof,Gogolin,Takahashi-book,Sachdev,phys-comb,mathphys}.

\par 
We explain in \S 2 some features  of 
the six-vertex model defined on a square lattice.  
We introduce  the Boltzmann weights and the transfer matrix 
for the six-vertex model. 
We review a method for diagonalizing the 
transfer matrix, which is called the coordinate Bethe ansatz,  
and we show the expressions of the free energy per site in 
the ferroelectric, the anti-ferroelectric, 
and the disordered phases, respectively 
 \cite{Lieb,Lieb-Wu}. 
The disordered phase is gapless, 
while the ferroelectric and the anti-ferroelectric phases 
have gaps \cite{Krinsky,Luther}. 
We derive a critical singularity appearing at the phase transition 
from the anti-ferroelectric to the disordered phases. 
 We review the calculation of the singular part of the free energy 
through the analytic continuation, as shown in Ref. \cite{Baxter-book}.  
The critical singularity is very weak and has the essential singularity 
 similar to the Kosterlitz-Thouless transition. 
We have thus derived the KT-like singularity 
through exact calculation. 
After  reviewing the finite-size analysis of conformal invariance 
\cite{Cardy,Bloete,Affleck,Affleck-review}, 
we discuss that the massless phase of the six-vertex model 
is related to 
 the conformal field theory of $c=1$ which has $U(1)$ symmetry.  
The $c=1$ CFT has the critical line where  critical 
 exponents change continuously with respect 
 to some parameter of the model \cite{Kadanoff,Cardy:c=1,SKYang}. 
There are quite a few papers on the finite-size corrections of the
integrable models \cite{deVega-Woynarovich,deVega-Karowski,Tsvelik,BIR,deVega-nested}. 
(For a review, see \cite{deVega,Korepin,Nagao}.) The critical line is 
 also characteristic to the Tomonaga-Luttinger liquid  \cite{Haldane}. 
 The existence of a critical line was first discovered by R.J. Baxter through 
 the exact solution of the eight-vertex model \cite{BaxterPRL}.

\par 
In \S 3, we review various integrable  models in statistical mechanics 
\cite{Baxter-book}. 
We briefly introduce the Ising model \cite{McCoy-Wu,Wu,Sato,Syoji}, 
the Potts models \cite{Potts,Kihara,TL,Inversion}, and the chiral Potts model 
\cite{Au-Yang,Baxter-Perk,Kino-McCoy,Kino-Perk,BS,Rittenberg,Sergeev,Perk-review,Fateev,Kashiwara-Miwa,Hasegawa-Yamada,Yamada}
and then the eight-vertex model \cite{BaxterPRL,Baxter72,Baxter73,Belavin} 
and IRF models \cite{ABF,Fusion-IRF,A-IRF,ABCD,Pearce,Kuniba,Akutsu,Pasquier}.  In \S 4, we solve explicitly the Yang-Baxter equations for the 
six-vertex model. We introduce the algebraic Bethe ansatz 
\cite{8VABA,Korepin82,Faddeev-Takhtajan,Takhtajan,Korepin}. 
Here we show that the Yang-Baxter equation of the algebraic Bethe ansatz 
 can be expressed by graphs.  
In \S 5, we discuss some mathematical theories 
associated with integrable models such as the braid group 
\cite{AW,WDA} and 
the quantum groups \cite{Jimbo-review}.

\par 
There have been novel developments in the mathematical physics 
associated with the six-vertex model \cite{Jimbo-review,Jimbo-Miwa}. 
The integrable vertex models associated with various Lie algebras 
have been obtained, 
which generalize the six-vertex model \cite{Jimbo-vertex,Bazhanov}. 
The crystal basis of the quantum groups 
is derived from the mathematical analysis of the 
corner transfer matrix  which is fundamental for calculating 
the one-point functions of the vertex models and IRF models 
\cite{Kashiwara,Lusztig}. 
Through the $q$-vertex operators, correlation functions 
of the XXZ spin chain or the six vertex model 
are obtained \cite{Jimbo-Miwa}. 
The dynamical Yang-Baxter equation \cite{BBB} and 
the elliptic quantum groups \cite{Felder,FV,twisted8V}
have also been extensively discussed. 
In fact, we can derive the $R$ matrix of the eight-vertex model 
systematically from the elliptic quantum group through 
the twists \cite{twisted8V}.  
Furthermore, 
the correlation functions of the XXZ model 
calculated with the $q$-vertex operators have been rederived 
for large but finite chains 
through the algebraic Bethe ansatz with the Drinfeld twists 
\cite{Maillet,KMT}. Here we note that 
the $q$-vertex operator can be defined only 
on the infinite chain, while the algebraic Bethe ansatz 
with the Drinfeld twists can be applied to any finite chain. 
By taking the thermodynamic limit, it has been shown that 
the two approaches indeed give the same results.  
These papers indeed illustrate nontrivial physical applications 
of the Drinfeld twits.      
It is recently found that the symmetry of 
the six-vertex model is enhanced at some particular coupling constants: 
the  transfer matrix commutes with  the generators 
of the $sl_2$ loop algebra  for the six-vertex model 
at roots of unity \cite{loop}.

\par   Let us discuss some physical motivations 
for  the six-vertex model. The exact solution of the 
six-vertex model was originally introduced  
for studying the statistical mechanics 
of ferro-electrics such as the residual entropy and 
the ferroelectric transitions \cite{Lieb,Lieb-Wu}. 
However, it seems that 
the physical motivation of the six-vertex model 
for ferro-electricity  has decreased.  
On the other hand, there are many different physical applications of 
of the six-vertex model. Here we consider a few examples, the domain wall 
theory \cite{Pokrovskii-Talapov,denNijs},  
crystal growth \cite{Beijeren,Saam,UGJ,Kim,Abraham,decoratedRSOS} 
and the thermodynamics of the XXZ spin chain through 
the quantum transfer matrix 
\cite{M.Suzuki-B,M.Suzuki-Inoue,Koma,SAW,Andreas,Shiroishi}.      
The crystal growth on surfaces has been discussed 
by applying exact solutions of the six-vertex model 
\cite{Beijeren,Saam,UGJ,Kim,Abraham} and some extensions 
\cite{decoratedRSOS}. The free energy of the six vertex model 
gives the equilibrium crystal shapes \cite{Beijeren,Saam}.   
The finite-temperature  
 thermodynamics of the XXZ spin chain has been studied 
 extensively through the 
quantum transfer matrix, 
 which is a version of the 
inhomogeneous six-vertex transfer matrix 
\cite{M.Suzuki-Inoue,Koma,SAW,Andreas,Shiroishi}. 
The quantum transfer matrix is obtained by regarding 
 one direction of the square lattice 
 as  the imaginary time or  the inverse temperature \cite{M.Suzuki-B}.   
 There have been considerable 
 efforts to evaluate thermal quantities analytically or numerically. 
Several functional equations on the eigenvalues of the transfer matrix 
have been devised \cite{Andreas,Shiroishi}.  Finally, we remark that 
a universal relation between the dispersion curve 
and the ground-state correlation length in quantum spin chains 
are discussed by using the exact solutions of the vertex models 
\cite{Okunishi}.

\par 
In \S 2, we employ mainly the notation 
of Baxter's textbook \cite{Baxter-book} except for the transfer matrix. 
In \S 4, however, we briefly show how the notation in statistical mechanics 
is related to the  notation of the algebraic Bethe ansatz 
or the Quantum Inverse Scattering Method.  
The graphical illustration should be useful. 
 Finally in \S 5 we discuss 
many connections of the six-vertex model 
 to several mathematical developments such as the quantum groups.

%
%

\setcounter{equation}{0} 
\renewcommand{\theequation}{2.\arabic{equation}}
\renewcommand{\thefigure}{2.\arabic{figure}}
\setcounter{section}{1}

\section{Solvable vertex models} 
\subsection{The six-vertex model}
\subsubsection{Ice rule}

\par 
Let us consider a square lattice as a model of 
 two-dimensional ferroelectric crystal. 
Molecules are placed on the vertices of the lattice. 
Arrows are placed on the edges of the lattice,  
which correspond to directions of dipole moments of hydrogen bonds.  
 As a crystal of hydrogen bonding, we may consider  ice, i.e.,  
 the crystal of water molecules.    
In this review, however, we simplify the molecular background 
of the model (For instance, see \cite{Ziman}). 
We assume that the dipole moments defined on  edges 
take only two values  $\pm 1$.

\par 
At a vertex in the lattice, there are four edges.  
There are possible 16 configurations of the four edges   
around the vertex since each of the edges 
takes two values $\pm 1$.

%
%
\begin{figure}[htb]
%
\begin{center}
\setlength{\unitlength}{1mm}
\begin{picture}(160,40)(0,0)
\put(80,20){\thicklines\line(-1,0){20}}
\put(80,20){\thicklines\vector(1,0){20}}
\put(80,20){\thicklines\vector(0,1){20}}
\put(80,0){\thicklines\line(0,1){20}}
\put(60,20){\makebox(5,5){$\alpha$}}
\put(80,0){\makebox(5,5){$\beta$}}
\put(80,35){\makebox(5,5){$\gamma$}}
\put(95,20){\makebox(5,5){$\delta$}}
\end{picture}
\end{center}
\caption{Configuration of polarizations 
around a vertex: $\al, \bt, \gm, \dt$.  
The Boltzmann weight is expressed  by 
$w(\alpha,\beta | \gamma, \delta)$.  
Positive directions of dipole moments are given by 
 upward or rightward.  }
\end{figure}
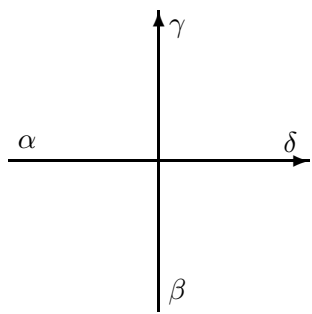

\par 
Let symbols  $\al, \bt, \gm, \dt$ denote 
the values of dipole moments around the vertex. 
Due to charge neutrality, 
they should satisfy the following condition
\be
\al+\bt = \gamma+ \dt 
\label{ice} 
\ee
For an illustration, let us consider the case when 
 $\al, \bt, \gm$ and $\dt$ are given by  +1.  
Then, $\al$  and $\bt$ give  + 2  to the vertex, while 
$\gm$ and $\dt$ deprive  + 2 from it, so that 
the net charge around the vertex is kept neutral: 
$\al + \bt - \gm - \dt=0$.

\par 
There are only 6 configurations satisfying the condition.   
The other configurations  that do not satisfy 
 the condition are not allowed   
 in thermal equilibrium. 
Thus, we call the model 
the six-vertex model. 
The condition (\ref{ice}) is sometimes called {\it ice rule}, 
since  ice as a crystal consists of water molecules connected 
by hydrogen bonding.

\par 
We  denote by 1 and 2 the values of polarization 
1 and $-1$, respectively.  
The symbols 1 and 2  are useful for  matrix notation. 
Let $p$ denote the notation of $\pm 1$  
and $k$ 1 and 2. Then, they are related by   
the relation: $k = 1+ (1-p)/2$.

%
%
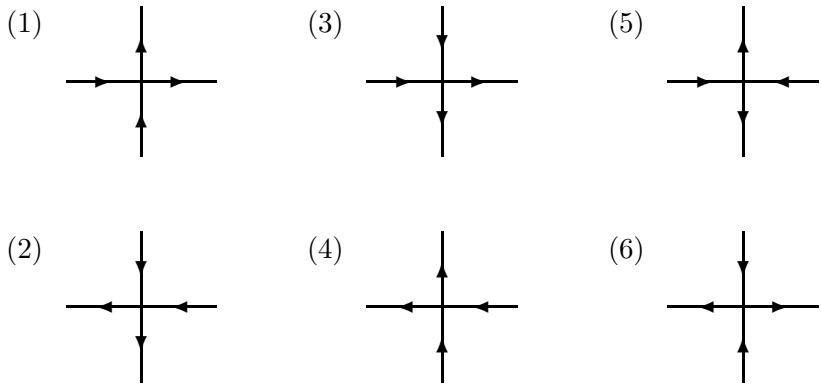
\begin{figure}[htb]
\begin{center}
\setlength{\unitlength}{1mm}
\begin{picture}(180,50)(-10,0)
\multiput(30,40)(40,0){3}{{\thicklines\line(-1,0){10}}}
\multiput(40,40)(40,0){3}{{\thicklines\line(-1,0){10}}}
\multiput(30,40)(40,0){3}{{\thicklines\line(0,1){10}}}
\multiput(30,30)(40,0){3}{{\thicklines\line(0,1){10}}}
%
\put(30,40){\thicklines\vector(1,0){6}} 
\put(20,40){\thicklines\vector(1,0){6}} 
\put(30,40){\thicklines\vector(0,1){6}} 
\put(30,30){\thicklines\vector(0,1){6}} 
%
\put(60,40){\thicklines\vector(1,0){6}} 
\put(70,40){\thicklines\vector(1,0){6}} 
\put(70,50){\thicklines\vector(0,-1){6}} 
\put(70,40){\thicklines\vector(0,-1){6}} 
%
\put(100,40){\thicklines\vector(1,0){6}} 
\put(120,40){\thicklines\vector(-1,0){6}} 
\put(110,40){\thicklines\vector(0,1){6}} 
\put(110,40){\thicklines\vector(0,-1){6}} 
\multiput(30,10)(40,0){3}{{\thicklines\line(-1,0){10}}}
\multiput(40,10)(40,0){3}{{\thicklines\line(-1,0){10}}}
\multiput(30,10)(40,0){3}{{\thicklines\line(0,1){10}}}
\multiput(30,0)(40,0){3}{{\thicklines\line(0,1){10}}}
%
\put(30,10){\thicklines\vector(-1,0){6}} 
\put(40,10){\thicklines\vector(-1,0){6}} 
\put(30,20){\thicklines\vector(0,-1){6}} 
\put(30,10){\thicklines\vector(0,-1){6}} 
%
\put(70,10){\thicklines\vector(-1,0){6}} 
\put(80,10){\thicklines\vector(-1,0){6}} 
\put(70,0){\thicklines\vector(0,1){6}} 
\put(70,10){\thicklines\vector(0,1){6}} 
%
\put(110,10){\thicklines\vector(-1,0){6}} 
\put(110,10){\thicklines\vector(1,0){6}} 
\put(110,20){\thicklines\vector(0,-1){6}} 
\put(110,0){\thicklines\vector(0,1){6}} 
\put(12,45){\makebox(5,5){(1)}}
\put(12,15){\makebox(5,5){(2)}}
\put(52,45){\makebox(5,5){(3)}}
\put(52,15){\makebox(5,5){(4)}}
\put(92,45){\makebox(5,5){(5)}}
\put(92,15){\makebox(5,5){(6)}}
\end{picture}
\end{center}
\caption{Vertex configurations satisfying the ice rule. They have the   
Boltzmann weights $w(\al,\bt | \gm, \dt)$ as follows:    
(1) $w(1,1|1,1)$; (2) $w(2,2|2,2)$; (3) $w(1,2|2,1)$;  
(4) $w(2,1|1,2)$; (5) $w(1,2|1,2)$; (6) $w(2,1|2,1)$.
The configurations (1) and (2) are for 
the weight $a$, (3) and (4) for $b$, and (5) and (6) for $c$.   
}
\end{figure}


\subsubsection{Boltzmann weights}

\par  It is a key idea in exactly solvable models 
 that we define the model  by the Boltzmann weights 
not by the energies of configurations. 
Let us introduce the Boltzmann weights for configurations 
around a vertex. For a vertex configuration $\al,\bt,\gm,\dt$, 
we denote by $\epsilon(\al,\bt | \gm,\dt)$ the energy at the vertex. 
Then, the Boltzmann weight for a temperature $T$ 
is given by 
\be
w(\al,\bt | \gm,\dt) = \exp(-\epsilon(\al,\bt | \gm,\dt)/k_B T) 
\ee
Under the ice rule, there are only six configurations allowed round a vertex. 
Here, it is assumed that the energy of a configuration violating the ice rule 
should be infinite. 
We denote by $\epsilon_j$ the energy of the $j$th vertex configuration 
shown in Fig. 2.2.  

\par 
Under no external field, the Boltzmann weights  must be invariant 
when reversing all the polarizations simultaneously. Thus, 
we have $\epsilon_1= \epsilon_2$, $\epsilon_3= \epsilon_4$ and    
$\epsilon_5= \epsilon_6$, when there is no  external field.   
We denote the Boltzmann weights as follows.    
\bea 
w(1, 1 | 1, 1) & = & w(2, 2 | 2, 2) = w_1 = a   \non \\ 
w(1, 2 | 2, 1) & = & w(2, 1 | 1, 2) = w_2 = b \non \\
w(1, 2 | 1, 2) & = & w(2, 1 | 2, 1) = w_3 =c 
\label{symmetry}
\eea

\par 
The Boltzmann weights of the zero-field six-vertex model 
have essentially only two parameters. For instance, we may 
choose $a/c$ and $b/c$. 
Note that the probability for the vertex configuration of $a$ 
is given by $a/(a+b+c)$, which does not change by
 replacing $a, b$ and $c$ with $\rho a$, $\rho b$ and $\rho c$.  
 
\par 
Let us consider $\pi/2$ rotation of the square lattice. 
If we rotate vertex configuration (1) of Fig. 2.2 
by the angle $\pi/2$ in the counterclockwise direction, 
then, it becomes vertex configuration (4). 
Under the $\pi/2$ rotation, the weight $a$ 
is exchanged with the weight $b$, 
while the weight $c$ does not change.

\subsection{Partition function and the transfer matrix}
\par 
Let us discuss the partition function of the system. 
We now set the  boundary conditions. Here, we consider 
the periodic boundary conditions for the two-dimensional lattice. 
We take a product of the Boltzmann weights over all the vertices 
of the lattice,  and sum up the product over all the 
allowed configurations of arrows on the lattice. 
\be
Z = \sum_{config} \prod_{j: vertices} w(a_j,b_j | c_j, d_j) 
\ee

\par 
The partition function of the square lattice can be formulated as 
the trace of the products of the transfer matrices. 
Let us define the transfer matrix  $\tau$ of the six-vertex model. 
The matrix elements of the transfer matrix  $\tau$ acting 
on $N$ lattice sites are given by 
\be
\tau^{a_1, \ldots, a_N}_{b_1, \ldots, b_N} 
= \sum_{c_1, \ldots, c_N} w(c_1, b_1 |  a_1, c_2)  w(c_2, b_2 | a_2, c_3 ) 
\cdots w(c_N, b_N | a_N, c_1 )  
\ee
%
%
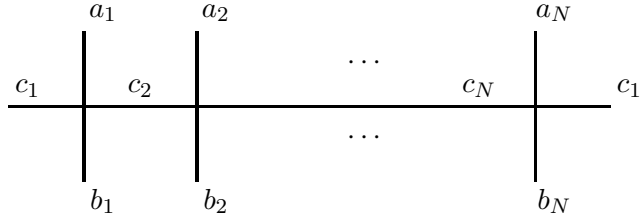
\begin{figure}[htb]
\begin{center}
\setlength{\unitlength}{1mm}
\begin{picture}(160,20)(-30,10)
\put(10,20){\thicklines\line(1,0){80}}
\multiput(20,20)(15,0){2}{{\thicklines\line(0,1){10}}}
\multiput(20,20)(15,0){2}{{\thicklines\line(0,-1){10}}}
\put(80,20){\thicklines\line(0,1){10}}
\put(80,20){\thicklines\line(0,-1){10}}
\put(10,20){\makebox(5,5){$c_1$}}  
\put(25,20){\makebox(5,5){$c_2$}} 
\put(70,20){\makebox(5,5){$c_{N}$}} 
\put(90,20){\makebox(5,5){$c_1$}} 
\put(55,25){\makebox{$\cdots$}}
\put(55,15){\makebox{$\cdots$}}
\put(20,30){\makebox(5,5){$a_1$}}  
\put(20,5){\makebox(5,5){$b_1$}} 
\put(35,30){\makebox(5,5){$a_2$}}  
\put(35,5){\makebox(5,5){$b_2$}} 
\put(80,30){\makebox(5,5){$a_N$}}  
\put(80,5){\makebox(5,5){$b_N$}} 
\end{picture} 
\end{center}
\caption{Matrix elements of the transfer matrix 
$\tau^{a_1, \ldots, a_N}_{b_1,\ldots, b_N}$} 
\end{figure}

\par 
Under the periodic boundary conditions, the partition function 
$Z_{NN^{'}}$ of $N \times N^{'}$ lattice 
is given by the trace of the $N^{'}$th power of the transfer matrices: 
\be
Z_{NN^{'}} =  {\rm Tr}\left( \tau^{N^{'}} \right) 
= \sum_{a_1, \ldots, a_N} \left( \tau^{N^{'}} 
\right)^{a_1, \ldots, a_N}_{a_1, \ldots, a_N} 
= \Lambda_1^{N^{'}} + \Lambda_2^{N^{'}} + 
\cdots + \Lambda_{2^N}^{N^{'}} 
\ee
Here $\Lambda_j$ denotes the eigenvalue of the transfer matrix $\tau$.

\par 
The free energy per site $f$ is given by 
\be 
f = - k_B T \log Z_{NN^{'}} /( N N^{'}) \, . 
\ee
In the thermodynamic limit: $N, N^{'} \rightarrow \infty$, 
the free energy per site is given by the 
largest eigenvalue $\Lambda_{max}$ of the transfer matrix $\tau$.

\subsection{Diagonalization of the transfer matrix }

\subsubsection{The Yang-Baxter relations for six-vertex model}

\par 
Let us consider three sets of the Boltzmann weights:  
$(w_1,w_2,w_3)=(a,b,c)$, $(a^{'}, b^{'}, c^{'})$, and 
$(a^{''}, b^{''}, c^{''})$. 
We denote by $\tau^{'}$ and $\tau^{''}$ the transfer matrices 
constructed from the sets of the Boltzmann weights 
 $(a^{'}, b^{'}, c^{'})$ and $(a^{''}, b^{''}, c^{''})$, respectively.  
If the three sets of the Boltzmann weights 
satisfy the Yang-Baxter equation 
\bea 
& & \sum_{\al,\bt,\gm} w(\al, \gm | a_1, a_2)
w^{'}(\bt, b_3 | \gm, a_3) w^{''}(b_1, b_2 | \al, \bt ) \non \\
& = & \sum_{\al, \bt, \gm} w^{''}(\bt, \al | a_2, a_3 ) 
w^{'}(b_1, \gm | a_1, \bt) w(b_2, b_3 | \gm, \al) \, , \label{ybr0} 
\eea
then the transfer matrices $\tau^{'}$ and $\tau^{''}$ commute. 
The derivation of the commutation relation is given in Appendix A. 
 We note that 
a graphical presentation of the Yang-Baxter equation (\ref{ybr0}) 
will be shown in Fig. 4.1.  

\par Let us define the parameter $\Delta$ as follows
 \be 
 \Delta = {\frac {a^2 +b^2 -c^2} {2ab}} \, . 
 \ee 
For the zero-field six-vertex model, we can show that  
if the two sets of the Boltzmann weights 
 have the same value of the parameter $\Delta$,  
 then their transfer matrices commute.  
We shall explicitly discuss in \S 4 
that it is indeed derived     
 from  the Yang-Baxter equations (\ref{ybr0}).

\subsubsection{The coordinate Bethe ansatz}

Let us consider the matrix element 
$\tau^{a_1, \cdots, a_N}_{b_1, \cdots b_N}$ 
of the transfer matrix $\tau$. 
Due to the ice rule, we may express the 
suffix $a_1, \ldots,a_N$ by the positions of the value 2, as follows. 
Suppose that there are $n$ suffices given by 
 the value 2 among the $N$ suffices $a_1, \ldots, a_N$. 
 The $n$ suffices are expressed as   
 $a_{x_1}, a_{x_2}, \ldots, a_{x_n}$ where $x_j$'s are in 
 increasing order: $x_1 < x_2 < \cdots < x_n$ .  
 Then, the entry $a_1, \ldots, a_N$ is equivalent to the set of  
  the $x_j$'s:  $x_1, \ldots, x_n$.  
For an illustration, let us consider the case $N=5$ and $n=3$. 
Then, $(x_1, x_2, x_3)=(1, 3, 4)$ corresponds to 
$(a_1, a_2, a_3, a_4, a_5) = (2, 1, 2, 2, 1)$. 
Thus, the matrix element 
 $\tau^{a_1 \cdots a_N}_{b_1, \ldots, b_N}$ can be denoted briefly  
as $\tau^{x_1, \ldots, x_n}_{y_1, \ldots, y_n}$. 

\par 
Let us now discuss how to solve  the secular equation:  
$\tau g = \Lambda g$.  Here, the transfer matrix 
$\tau$ is a $2^N \times 2^N$ matrix,  
 $g$ is a  $2^N$-dimensional eigenvector with  eigenvalue $\Lambda$.  
In terms of matrix elements, the secular equation is written as     
\be 
\sum_{y_1, \ldots, y_n} 
\tau^{x_1, \ldots, x_n}_{y_1, \ldots, y_n} 
g(y_1, \ldots, y_n) 
=   \Lambda \, g(x_1, \ldots, x_n)
\ee
Here, $g(x_1, \ldots, x_n)$ denotes the matrix element of 
the vector $g$ for the entry $(x_1, \ldots, x_n)$. 
In the coordinate Bethe ansatz, 
we assume the following form for the matrix element 
of the possible eigenvector $g$ 
\be 
g(x_1, \ldots, x_n) = \sum_{P \in S_n} A_P \exp\left( 
k_{P1} x_1 + \cdots + k_{Pn} x_n \right)
\label{g-BA}
\ee
Here, $S_n$ denotes the symmetric group of order  $n$, 
and $P$ is a permutation of $n$ letters, $1, 2, \ldots, n$, 
where  $P$ maps $j$ into  $Pj$.   
The expression (\ref{g-BA}) is called the Bethe ansatz wave-function. 
If the vector $g$ whose elements are of the form (\ref{g-BA}) 
is an eigenvector of the transfer matrix, 
 then we call it a Bethe ansatz eigenvector.

\par 
For general $n$, the vector (\ref{g-BA}) becomes  
an eigenvector of the transfer matrix, if the wave-numbers $k_j$'s 
satisfy the Bethe ansatz equations. They are given  by the following 
\be 
\exp(i N k_j ) = (-1)^{n-1} \prod_{\ell=1}^{n} \exp(-i \Theta(k_j, k_{\ell})) 
\, , \quad {\rm for} \quad j=1, \ldots, n, 
\label{BAE}
\ee
 where $\Theta(p,q)$ is defined by 
 \be 
 \exp(-i \Theta(p, q)) = 
{ \frac {1 - 2 \Delta e^{i p} + e^{i(p+q)}}
       {1 - 2 \Delta e^{i q} + e^{i(p+q)}} }
 \label{phasefactor}
 \ee
 For the solutions $k_j$'s to the Bethe ansatz equations,  
the eigenvalue $\Lambda$ of the transfer matrix is given by 
\be 
\Lambda(k_1, \ldots, k_n) = a^N \, L(z_1) \cdots L(z_n) + 
b^N \, M(z_1) \cdots M(z_n) \, , \label{evBA}
\ee
where $z_j= \exp( i k_j)$ for $j=1, \ldots, n$ and the functions 
 $L(z)$ and $M(z)$ are defined by 
 \be 
 L(z) = {\frac {ab + (c^2-b^2)z} {a(a-bz)}} \quad , 
 \qquad 
 M(z)= {\frac {a^2- c^2 - abz} {b(a-bz)}} \, . 
\label{LM}
 \ee

\par 
When we discuss the spectrum of an integrable model 
through the coordinate Bethe ansatz,  
we often assume that all the eigenvectors of the transfer matrix 
are characterized by the Bethe ansatz wave-function (\ref{g-BA}).  
However, it is not certain whether the assumption is valid or not. 
Thus, we have to check it by other methods.  In fact,    
there are several numerical studies on   
 the validity of the completeness of the Bethe ansatz 
 for some integrable models.  

\par 
On the other hand, there is no doubt on the mathematical structure 
of the Bethe ansatz wave-function. 
We can derive the expression (\ref{g-BA})  
 by the algebraic Bethe ansatz through the `two-site' model 
 \cite{Korepin82,Korepin}.
 (There is an instructive notice in \cite{infinite}.) 
 It was shown   that the matrix elements of  the product of 
 $B$ operators acting on the vacuum 
are given by the Bethe ansatz wave-function (\ref{g-BA}) 
with $k_j$'s being generic.   

\par 
The Yang-Baxter relation leads to not only the 
integrability of the six-vertex model but also 
the systematic construction of the eigenvectors. 
In fact, we shall see in \S 4 that the algebraic Bethe ansatz 
is solely based on the Yang-Baxter equation.

\subsubsection{An example of the eigenvector}
\par 
For an illustration,  let us consider the eigenvector $g$ 
for the case of $n=1$:  $g(x) = A \exp(i k x)$. Through a 
direct calculation, we have 
\bea 
\sum_{y=1}^{N} \tau^{x}_{y} g(y) & = & 
\sum_{y=1}^{x-1} a^{x-y-1} b^{N-x+y-1} c^2 g(y)  
+\sum_{y=x+1}^{N} a^{N+x-y-1} b^{y-x-1} c^2 g(y)  \non \\
& = &  \left( a^N  \, L(z) + b^N \, M(z) \right) g(x) + 
 {\frac {a^{N-1} b^{N-x} c^2 z} {a- bz}} (1 -z^N) 
\eea
Therefore, $g(x)$ becomes an eigenvector if 
the Bethe ansatz equation $z^N=1$ is satisfied.

\subsection{The free energy of the six-vertex model }
\subsubsection{Three phases of the six-vertex model}

There are three phases for the zero-field six-vertex model. 
They are given by the regions of the parameter $\Delta$:  
ferroelectric phase when $\Delta > 1$; 
anti-ferroelectric phase when $\Delta < -1$; disordered phase 
when $-1 <\Delta < 1$.  It is found that the disordered phase
($-1 <\Delta < 1$)  
is gapless (massless), 
while the ferroelectric phase ($\Delta >1$)
 and the anti-ferroelectric phase ($\Delta <-1$) are gapful  (massive).

\par 
Let us consider the phase diagram of the six vertex model shown in Fig. 2.4.  
We recall that the ratios $a/c$ and $b/c$ determine the model.  
Here we note that the number of independent parameters is given by two, 
since the overall normalization factor is arbitrary.  
The regimes I and II give  the ferroelectric phase, 
the regimes III and IV are  anti-ferroelectric and disordered, 
respectively. In terms of the Boltzmann weights, the regime I is given by  
$a> b+c$,  the regime II by  
$b > a+c$, and the regime IV by  $a+b < c$ . Then, the regime III 
is given by  $a <b+c$, $b < c+a$ and $c < a +b$.

\par 
Let us derive the three phases through an intuitive argument. 
For the ferroelectric regime of $\Delta > 1$, 
we have $a^2 + b^2 - c^2 > 2 a b$ ,  
which leads to  the inequality:  $|a - b | > c$. 
When $a> b$, we have $a > b+c$. Thus, 
the configuration where all 
the Boltzmann weights are given by the weight $a$ should be 
the largest contribution to the partition function $Z$.   
In fact, when $n=0$ in eq. (\ref{evBA}), 
we have $\Lambda=a^N + b^N$. When $a> b$, 
the free energy per site, $f$,  is given by $\epsilon_1$: $f=-k_B T \log a$.  
When $b> a$, we have $f=\epsilon_3$ , similarly.  

\par 
For the anti-ferroelectric regime of $\Delta < - 1$, 
we have the inequality $a + b  < c$.  Thus, 
 the vertex configurations for $c$ 
should be more favorable than those of $a$ or $b$.  
In fact, it is  shown that the transfer matrix has  
the largest eigenvalue  when $n=N/2$. 
Furthermore, if 
 we send  $|\Delta|$  to infinity, 
all the vertex configurations should be  given by those of $c$.    
The phase is thus called anti-ferroelectric.  

\par 
We may explain the reason why it is called anti-ferroelectric. 
Let us consider the configuration (5) of Fig. 2.2.  
We see that  the arrow coming from the left goes upward, while 
the arrow coming from the right goes downward. 
If all the vertex configurations on the square lattice  
are given by (5), then the lines coming from South West to North East 
and the lines coming from North East to South West occupy 
the lattice alternatively.  
This gives the anti-ferroelectric order. 

%
%
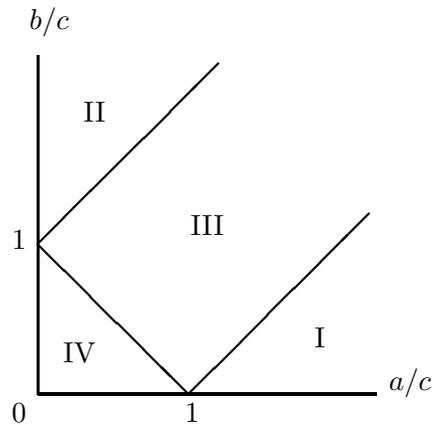
\begin{figure}[htb]
\begin{center}
\setlength{\unitlength}{1mm}
\begin{picture}(80,60)(0,-10)
\put(0,0){\thicklines\line(0,1){45}}
\put(0,0){\thicklines\line(1,0){45}}
\put(0,20){\thicklines\line(1,-1){20}}
\put(0,20){\thicklines\line(1,1){24}}
\put(20,0){\thicklines\line(1,1){24}}
\put(47,-1){\makebox(5,5){$a/c$}}  
\put(-1,47){\makebox(5,5){$b/c$}} 
\put(35,5){\makebox(5,5){I}}  
\put(5,35){\makebox(5,5){II}} 
\put(20,20){\makebox(5,5){III}}  
\put(3,3){\makebox(5,5){IV}} 
\put(18,-5){\makebox(5,5){1}} 
\put(-5,18){\makebox(5,5){1}} 
\put(-5,-5){\makebox(5,5){0}} 
\end{picture}
\caption{Phase diagram of the six-vertex model: regimes I and II are ferroelectric, 
regime III is disordered, and regime IV is anti-ferroelectric.}
\end{center}
\end{figure}

\subsubsection{Parametrization of the Boltzmann weights} 
 \par 
 It is nontrivial how to  parametrize the Boltzmann weights. 
Recall that there are two independent parameters 
 for the six-vertex model.  
 Thus, if we consider  $\Delta$ as a parameter, 
  there is only another one.  As we shall see later, 
 it is related to the spectral parameter. 

\par 
We recall that the phases of the zero-field six-vertex model 
are classified into the following: 
$\Delta < -1$, $-1 < \Delta < 1$ and $1 < \Delta$.

\par \noindent 
(1) {\it Anti-ferroelectric phase}
 \par 
For $\Delta < -1$, we define a real parameter $\lambda$ by 
 \be
 \Delta = -\cosh \lambda \qquad (0 < \lambda ) 
 \ee
and we parametrize the Boltzmann weights as follows 
\be
a  = \rho \sinh(\frac {\lambda-v} 2) \, , \quad
 b = \rho \sinh(\frac {\lambda + v} 2) \, ,  \quad  
c = \rho \sinh \lambda \, , \quad ( - \lambda < v < \lambda)     
\ee
where $\rho$ is the normalization factor. 
Let us define the rapidity $\alpha$ for the wavenumber $k$ as  
\be 
\exp(i k) = 
{\frac {e^{\lambda } - e^{- i \alpha}} {e^{\lambda - i \alpha} -1}} 
= - {\frac {\sin{\frac 1 2}(\alpha - i \lambda)} 
           {\sin{\frac 1 2}(\alpha + i \lambda)}} 
\ee
By replacing the wave numbers $p$ and $q$ 
with the rapidities $\alpha$ and $\beta$ in eq. (\ref{phasefactor}), 
 the phase factor $\Theta(p,q)$ is written  as    
follows 
\be 
\exp(- i \Theta(p,q)) = 
{\frac {e^{2 \lambda } - e^{-i (\alpha- \beta)} } 
      {e^{2 \lambda -i ( \alpha -\beta) } - 1}} 
      = - {\frac {\sin {\frac 1 2}((\alpha -\beta) - 2 i \lambda)}
              {\sin {\frac 1 2}((\alpha -\beta) + 2 i \lambda)}}
\ee

\par \noindent 
(2) {\it Disordered phase}
\par 
For $-1 < \Delta < 1$, we define a positive real parameter $\mu$ by 
 \be
 \Delta = -\cos \mu \qquad ( 0 < \mu < \pi )
 \ee
and we parameterize the Boltzmann weights as follows 
\be
a = \rho \sin(\frac {\mu-w} 2) \, , \quad 
b = \rho  \sin(\frac {\mu + w} 2) \, , \quad   
c = \rho \sin \mu  \, ,  \quad ( -\mu < w < \mu ) \, .   
\label{weightAF}
\ee
Here $\rho$ is the normalization factor. 
We define the rapidity $\alpha$ for the wavenumber $k$  by 
\be 
\exp(i k) = {\frac {e^{i \mu} - e^{\alpha}} {e^{i \mu + \alpha} -1}} 
= - {\frac {\sinh {\frac 1 2}(\alpha - i \mu)} 
           {\sinh {\frac 1 2}(\alpha + i \mu)} } 
\ee
In terms of rapidities $\alpha$ and $\beta$,  
the phase factor $\Theta(p,q)$ is expressed as  
\be 
\exp(- i \Theta(p,q)) = 
{\frac {e^{2 i \mu } - e^{\alpha- \beta}} 
      {e^{2 i \mu + \alpha -\beta } - 1}} 
      = - {\frac {\sinh {\frac 1 2}((\alpha -\beta) - 2 i \mu )}
              {\sinh {\frac 1 2}((\alpha -\beta) + 2 i \mu )}}
\ee

\par \noindent 
(3) {\it Ferroelectric phase}
\par  
For $\Delta > 1$, we define the real parameter $\lambda$ by 
 \be
 \Delta =  \cosh \lambda \qquad (0 < \lambda ) 
 \ee
and we may parameterize the Boltzmann weights as follows: 
When $a > b$ (the regime I), 
\be
a  =  \rho \sinh(\frac {\lambda-v} 2) \, , \quad
 b = - \rho \sinh(\frac {\lambda + v} 2) \, ,  \quad  
c = \rho \sinh(\lambda)  \, , \quad    
( v < -\lambda ) \, . 
\ee
When $a < b$ (the regime II),   
\be
a  = - \rho \sinh(\frac {\lambda-v} 2) \, , \quad
 b =  \rho \sinh(\frac {\lambda + v} 2) \, ,  \quad  
c = \rho \sinh(\lambda)  \, ,  \quad    
( v >  \lambda ) \, . 
\ee 
Here we recall that $\rho$ is the normalization factor.  
The wavenumber $k$ is related to the rapidity $\alpha$  by 
\be 
\exp(i k) = - {\frac {e^{\lambda } - e^{- i \alpha}} 
{e^{\lambda - i \alpha} -1}} 
= {\frac {\sin{\frac 1 2}(\alpha - i \lambda)} 
{\sin{\frac 1 2}(\alpha + i \lambda)}} 
\ee
The phase factor $\Theta(p,q)$ is written as     
\be 
\exp(- i \Theta(p,q)) = 
{\frac {e^{2 \lambda } - e^{-i (\alpha- \beta)} } 
      {e^{2 \lambda -i ( \alpha -\beta) } - 1}} 
      = - {\frac {\sin {\frac 1 2}((\alpha -\beta) - 2 i \lambda )}
                 {\sin {\frac 1 2}((\alpha -\beta) + 2 i \lambda )}}
\ee

\par 
Let us consider curved lines given by changing 
the spectral parameter $w$ or $v$ 
continuously. All the curves pass through the points 
$(a/c, b/c)$=$(1, 0)$ and $(0,1)$. Except for the two points, however,  
any points on the horizontal axis: $b/c=0$ or the vertical axis: $a/c=0$ 
 are never reached by the above parametrizations with  finite values.

\subsubsection{Expressions of the free energy}

\par \noindent 
(1) {\it Anti-ferroelectric phase}
 \par 
When $\Delta < -1$, the system is in the anti-ferroelectric phase. 
 The free energy per site $f$ is given by   
\bea 
f & = & - k_B T \log a - k_B T  \left( {\frac {\lambda +v} 2} + 
\sum_{m=1}^{\infty} {\frac {e^{-m \lambda} \sinh m(\lambda + v)}
                           { m \cosh m \lambda}} \right)  \\
\label{f_AF}
 & = & - k_B T \log b - k_B T  \left( {\frac {\lambda - v} 2} + 
\sum_{m=1}^{\infty} {\frac {e^{-m \lambda} \sinh m(\lambda - v)}
                           { m \cosh m \lambda}} \right)                            
\eea
for $- \lambda < v < \lambda$.

\par \noindent 
(2) {\it Disordered phase}
 \par 
When $-1 < \Delta < 1$, we have  
\bea
f & = & - k_B T \log a - k_B T 
\int_{-\infty}^{\infty}  
 {\frac {\sinh (\mu + w) x  \sinh (\pi - \mu)x } {2 x  \sinh \pi x \cosh \mu x }} dx
\label{f_DO}  \\
 & = & - k_B T \log b - k_B T 
\int_{-\infty}^{\infty}  
 {\frac {\sinh (\mu - w) x  \sinh (\pi - \mu)x } {2 x  \sinh \pi x \cosh \mu x }} dx
\eea
for $-\mu < w < \mu$.

\par \noindent 
(3) {\it Ferroelectric phase}

When $a>b$ (the regime I), we have  $f= - k_B T \log a$, 
while when $b> a$ (the regime II), we have $f= - k_B T \log b$. 

\subsubsection{Correlation length}

Correlation length has been calculated for the six-vertex model 
\cite{Krinsky,Baxter-book}. In fact, it is calculated for the
eight-vertex model. In the ferroelectric and the anti-ferroelectric regimes 
of the six-vertex model, the correlation length is finite. 
It becomes very large  near their phase boundaries to the disordered phase. 
In the disordered phase of the six-vertex model, the correlation length diverges 
throughout the regime. Thus, the disordered phase is critical.

\subsection{Critical singularity in the anti-ferro regime 
near the phase boundary}

We now discuss a singular behavior of the free energy shown 
at the phase transition from the anti-ferroelectric to the  
disordered phases \cite{Baxter-book}. 
Here we note that the former one is gapful, 
while the latter is gapless.  
Recall that the phase boundary between the regimes III and IV 
is given by $a+b=c$.  When $T > T_c$, the system should be 
 in the disordered regime ($ -1 < \Delta < 1)$, 
 while when $T < T_c$ it is in the anti-ferroelectric regime ($\Delta < -1)$.  
In the lower temperature, the system should be ordered.

\par 
Let us calculate the analytic continuation of the 
high-temperature free energy (\ref{f_DO}) 
into the low-temperature phase, so that  
we can single out the singularity of the free energy 
near $T_c$ and for $T < T_c$.  
First, we reformulate the integral of eq. (\ref{f_DO}) as follows. 
\be 
 \int_{-\infty}^{\infty} 
{\frac {\sinh (\mu+w)x \sinh (\pi- \mu)x }   
      { 2 x \sinh \pi x \cosh \mu x}} dx = 
{\cal P} \int_{-\infty}^{\infty} 
{\frac {\sinh(\mu+w)x \exp (\pi- \mu)x }   
      { 2 x \sinh \pi x \cosh \mu x}} dx 
\ee
Here ${\cal P}$ denote the principal value integral. 
Second,  we set  $\lambda$ to be a very small positive real number, 
and take a value  $v$ satisfying  $-\lambda < v < \lambda$.  
We consider the path in the complex $\mu$-plane: 
\be 
\mu= \lambda \exp( -i \theta)\, , \quad {\rm for } \quad 
0 \le \theta \le \pi/2 \, . 
\ee
Along the path, we  calculate the analytic continuation of 
the high-temperature free energy (\ref{f_DO}). 
Here, we also consider the path of $w$: 
$w= v \exp(-i \theta)$ for  $0 \le \theta \le \pi/2$. 
Then, we have 
\bea 
& & {\cal P} \int_{-\infty}^{\infty} 
{\frac {- i \sin(\lambda + v)x \exp(\pi + i  \lambda )x }   
      { 2 x \sinh \pi x \cos \lambda x}} dx \non \\
&=  & {\frac {\lambda + v} 2} + \sum_{m=1}^{\infty} 
{\frac {e^{-m \lambda} \sinh m(\lambda+v)}
       {m \cosh m \lambda} } 
        - i \sum_{m=1}^{\infty} 
{\frac {(-1)^m \cos\left((m-1/2)\pi v /\lambda \right) 
 e^{-(m-1/2) \pi^2/\lambda} }   
  {(m-1/2) \sinh \left((m-1/2)\pi^2/\lambda \right)}}
\eea   
The real part of the analytic continuation corresponds to 
the expression of the anti-ferroelectric free energy. 
Therefore, we  obtain the singular part of the free energy 
\be 
f_{sing} = i k_B T \sum_{m=1}^{\infty} 
{\frac {(-1)^m \cos\left((m-1/2)\pi v /\lambda \right) 
\exp \left( -(m-1/2) \pi^2/\lambda \right)} 
  {(m-1/2) \sinh \left((m-1/2)\pi^2/\lambda \right)}}
\ee

\par 
We define the reduced temperature $t$ by 
\be 
t = (a+b-c)/c \, . 
\ee
In the low-temperature phase ($T < T_c$) and near $T_c$, 
$t$ is given by 
\be 
t \approx - {\frac 1 8}(\lambda^2 -v^2)
\ee
Approximately,  $t$ is given by $t \approx - \lambda^2/8 $. 
Near $T_c$, we have 
\be 
f_{sing} \approx - 4 i k_B T e^{- \pi^2/\lambda} \cos({\frac {\pi v} {2 \lambda}})
\ee
Thus, we have 
\be 
f_{sing} \propto \exp \left( - {\frac {\rm constant} {\sqrt{-t}}  }  \right)
\ee
Near $T_c$, the free energy has  an essential singularity.

\par 
The singularity of the free energy is very close to 
that of the Kosterlitz-Thouless transition. In fact, 
calculating exactly, we can show that 
the correlation length $\xi$ diverges 
at $T_c$ as $\xi \propto \exp \left( {\rm constant}/{\sqrt{-t}} \right)$, 
when  $T$ approaches $T_c$  in the anti-ferroelectric phase.

\subsection{XXZ spin chain and the transfer matrix}

The logarithmic derivative of the transfer matrix   of the 
six-vertex model gives the Hamiltonian of the XXZ spin chain. 
\be
{\frac d {d v}} \log \tau |_{v=-\lambda} = \tau^{-1} 
{\frac d {d v}} \tau  \propto H_{XXZ} + constant
\label{logdv}
\ee
where $H_{XXZ}$ is given by 
\be 
H_{XXZ} =  J \sum_{j=1}^{L} \left(\sigma_j^X \sigma_{j+1}^X +
 \sigma_j^Y \sigma_{j+1}^Y + \Delta \sigma_j^Z \sigma_{j+1}^Z  \right) \, . 
\label{hxxz}
\ee
Intuitively, we may express it by $\tau_{6V}(v) \approx \exp(- v H_{XXZ})$. 

\par 
The XXZ spin chain and the six-vertex transfer matrix 
have the same eigenvectors in common thanks to eq. (\ref{logdv}).  
Taking logarithm of the Bethe ansatz equations, we have 
\be 
N k_j = 2\pi I_j - \sum_{\ell=1}^{M} \Theta(k_j, k_{\ell}) \, , \quad {\rm for } \quad 
j = 1, \ldots, M\, .  
\ee
where $M$ is the number of down spins. (We assume  $2M \le N$.)  
Here $I_j$ is an integer if $M$ is odd and half an integer if $M$ is even.  

\par 
The ground state of the XXZ spin chain for $\Delta <1$ was obtained by 
Yang and Yang \cite{Yang-Yang}.
 The ground state  is specified by the integers 
$I_j= j - (M+1)/2$ for $j=1, \ldots, M$. 
When $N$ is very large, 
the distribution of $k_j$'s becomes continuous.  
The number of $k_j$'s between 
$k$ and $k + dk$ can be approximated by  $N \rho(k) dk$.  
Thus, we have the integral equation of $\rho(k)$ 
\be
2 \pi \rho(k) = 1 + \int^{Q}_{-Q} {\frac {\partial \Theta(k,k^{'})} {\partial k}} 
\rho(k^{'}) dk^{'}
\label{intQ}
\ee
where $Q$ is determined by the normalization condition 
\be 
\int^{Q}_{-Q} \rho(k) dk = M/N \, . 
\ee
The integral equation (\ref{intQ}) can be solved by 
changing the variable $k$ to rapidity $\alpha$, 
and then by taking the Fourier transform for the half-filling case $M/N=1/2$. 
When $M/N$ is close to $1/2$, the integral equation can be solved 
by the Wiener-Hopf method \cite{Yang-Yang}.

\subsection{Low-lying excited spectrum of the transfer matrix 
and  conformal field theory}

In this section,  
we assume that the low excited spectra of the transfer matrix of 
gapless models should be characterized by the conformal invariance, 
if the system size is large enough. 
The assumption is not rigorous, however, 
there are many studies which confirm it numerically for 
integrable models. We review the finite-size corrections  
for the XXZ spin chain and  the six-vertex model. 

\subsubsection{Finite-size corrections}

Let us consider a conformally  invariant field theory 
defined in  the two-dimensional 
Euclidean space with coordinates $r_1$ and $r_2$. 
The energy momentum tensor $T_{\mu \nu}$ for $\mu, \nu =1, 2$ 
should be symmetric and traceless due to the conformal symmetry. 
Introducing the complex coordinates 
 $z= r_1 + i r_2$,  ${\bar z} = r_1 - i r_2$, 
 we define the chiral operator : $T= (T_{11}- T_{22} - 2i T_{12})/4$,  
 and the anti-chiral operator: ${\bar T}= (T_{11}- T_{22} + 2i T_{12})/4$. 
The operator $T$ (or ${\bar T}$) depends only on the variable 
$z$ (or ${\bar z}$).

\par 
The energy momentum tensor has the operator product expansion
\be 
T(z_1)T(z_2) = {\frac {c/2} {(z_1-z_2)^4}} + {\frac {2 T(z_2)} {(z_1-z_2)^2}} 
+ {\frac {\partial T(z_2)} {z_1-z_2}} + \cdots 
\label{OPE}
\ee 
Here $c$ is called the central charge. 
We define  the operators $L_n$ by the expansion:   
$T(z) = \sum_{n=-\infty}^{\infty} L_n z^{-n-2}$. 
The operator product expansion (\ref{OPE})
corresponds to  the Virasoro algebra 
\be 
[L_n, L_m] = (m-n) L_{m+n} + {\frac c {12}} (m^3-m) \delta_{m+n, 0} 
\ee
Under a conformal transformation $z \rightarrow w$, 
the energy momentum tensor is transformed as 
\be 
T(z) = \left( {\frac {dw} {dz}} \right)^2 {\tilde T}(w) + {\frac c {12}} 
\{w, z \}
\label{transfT}
\ee
where the symbol $\{w, z \}$ denotes the Schwarzian derivative:  
$(d^3 w/dz^3)/(dw/dz)$ - $(3/2) (d^2 w/dz^2)^{2}/(dw/dz)^2$. 

\par 
Let us consider the conformal mapping from the $z$-plane to 
a cylinder of circumference $L$: 
\be 
z \rightarrow w = {\frac L {2 \pi}} \log z  
\ee
Here $w= \tau - ix$ with imaginary time $\tau = i t$. 
The Hamiltonian ${\hat H}$ on the cylinder 
is given by the space integral of 
the (1,1) component of the energy momentum tensor $(T_{cyl})_{\mu \nu}$ 
\bea 
{\hat H} & = & {\frac 1 {2 \pi}} 
\int_{0}^{L} dx \left( T_{cyl}(w) + {\bar T}_{cyl}(w) \right) \non \\
 & = &  {{2 \pi} \over L}(L_0 + {\bar L}_0) - {\frac {\pi c } {6 L}} 
\label{finiteE}
\eea
Here we have used (\ref{transfT}). For the momentum operator 
on the cylinder, we have 
\bea 
{\hat P} & = & {\frac 1 {2\pi}} 
\int_{0}^{L} dx \left( T_{cyl}(w) - {\bar T}_{cyl}(w) \right) \non \\
& = & {\frac {2 \pi} L} (L_0 - {\bar L}_0) 
\label{finiteP}
\eea

\par 
Let us now discuss the application of the formulas (\ref{finiteE}) and (\ref{finiteP}) 
to the quantum spin chains. We assume that the low excited energies 
 should be gapless and conformally  invariant. In other words, we  assume  
that  the excitations near the ground state have 
a linear dispersion relation. Let $v$ denote the velocity of the 
linear dispersion.  Then,  for the ground-state energy $E_0$,  we have 
\be 
E_0 = L e_{\infty} - {\frac {\pi v c} {6L}}  
\ee
and for the excited energy $E_{ex}$ and the momentum $P_{ex}$ we have 
\bea
E_{ex} - E_0 & = & {\frac {2 \pi v} L}(h + {\bar h}+ {N}+ {\bar N}) \non \\ 
P_{ex} - P_0  & = & {{2 \pi } \over L}(h - {\bar h} + N- {\bar N}) 
\eea
where $h$ and ${\bar h}$ are the conformal weights related to 
the zero modes of the field, and the eigenvalues of 
$N$ and ${\bar N}$ are given by non-negative integers. 

\par 
There is another viewpoint on the finite-size scaling. 
Let us consider the $t$ axis as the space axis for 
an infinitely long  quantum spin chain, 
and $x$ axis as the imaginary time axis.  
Here we assume that $L= v \beta = v/T$. 
 Thus, our system now becomes the quantum spin chain in the finite 
 temperature $T$. Replacing $E_0$ with $v \beta f$, where 
 $f$ denotes the finite-temperature free energy of the spin chain, 
 we have $f = - \pi c T^2 /6v$. Thus, we may calculate 
 the specific heat $C$  by the formula $C = -T \partial^2 f/\partial T^2$,  
 and we have  
\be 
C = {\frac {\pi c } {3 v}} T
\ee

\subsubsection{The free Boson: CFT with $c=1$}

\par 
Let us consider a free Bose field  $\varphi(x,t)$ defined on 
a cylinder of circumference $L$. 
The Lagrangian  is given by 
\be 
{\cal L} = {\frac 1 2}\, g \int dx \left\{ \left(\partial_t \varphi \right)^2
- \left(\partial_x \varphi \right)^2 \right\} 
\ee 
We define  the mode $\varphi_n$ by the Fourier expansion  
$\varphi(x,t) = \sum \varphi_n(t) \exp(-2\pi i n x/L) $ . 
 From the canonical quantization,  
we have the conjugate momentum $\pi_n = gL {\dot \varphi}_{-n}$, 
and the commutation relation $[\varphi_n, \pi_m] = i \delta_{nm}$ . 
With the operators $a_n$ and ${\bar a}_n$ for $n \ne 0$ satisfying 
\be 
[a_n , a_m] = n \delta_{n+m}  \quad 
[a_n , {\bar a}_m] = 0  \quad [{\bar a}_n , {\bar a}_m] = n \delta_{n+m}  \, , 
\ee
the Fourier mode is expressed as 
$\varphi_n = i  \left(a_n - {\bar a}_{-n} \right)/(n {\sqrt{4 \pi g}})$ ,  
for  $n \ne 0$. The Hamiltonian is given by 
\be 
{\cal H} = {\frac 1 {2gL}} \pi_0^2 + {\frac {2 \pi} {2 L}} 
\sum_{n \ne 0} (a_{-n} a_n + {\bar a}_{-n} {\bar a}_n ) 
\ee
Hereafter, we assume $g=1/4\pi$.  The convention is consistent with the 
conformally invariant partition functions.

\par 
We now discuss the compactification of the Boson with radius $R$. 
Suppose that  the  field operator $\varphi$ takes its  
value only on the circle of radius $R$. In other words,  
we may identify $\varphi$ with $\varphi + 2 \pi R$. 
Then, the eigenvalue of the momentum $\pi_0$ conjugate to $\varphi_0$ 
is given by $n/R$ for an integer $n$. Here we recall that 
the wavenumber of a one-dimensional system of size $L$ 
is given by $2 \pi n/L$ ($n \in {\bf Z}$), and also that 
 the range of $\varphi_0$ is given by $2 \pi R$, which corresponds to $L$. 
Furthermore, we may assign on the operator $\varphi$ 
 the boundary condition for an integer $m$ 
\be 
\varphi(x+L, t) = \varphi(x,t) + 2 \pi  m R 
\ee 
Here, the integer $m$ is called the {\it winding number}. 
The mode expansion of $\varphi$ is  given by 
\be 
\varphi(x,t) = \varphi_0 + {\frac {4 \pi} L} \pi_0 t + 
{\frac {2 \pi R m} L} x  
+  i  \sum_{n \ne 0} {\frac 1 n} \left( 
a_n e^{2 \pi i n (x-t)/L} 
- {\bar a}_{-n} e^{2 \pi i n (x + t )/L}  \right)
\ee
In terms of the coordinates $z= \exp(2 \pi(\tau - i x)/L)$ 
and ${\bar z} = \exp(2 \pi(\tau + i x)/L)$, $\varphi(x, t)$ is given by 
the sum of the holomorphic and anti-holomorphic parts:    
$\varphi(z,  {\bar z}) = \phi(z) + {\bar \phi}({\bar z})$ . 
Here, they are  given by 
\be 
\phi(z) = {\frac {\varphi_0} 2} - i a_0 \log(z) + i \sum_{k \ne 0} 
{\frac 1 k} a_k z^{-k} 
\quad 
{\bar \phi}({\bar z}) = {\frac {\varphi_0} 2} - i {\bar a}_0 \log({\bar z}) 
+ i \sum_{k \ne 0} {\frac 1 k} {\bar a}_k {\bar z}^{-k} 
\ee
with  $a_0= n/R + mR/2$ and   ${\bar a}_0= n/R - mR/2$ .
Here we can show that the operator $J(z)= i \partial \phi(z)/\partial z$
is the $U(1)$ current operator.

\par 
Making use of Noether's theorem, we have 
\be 
T(z) = - {\frac 1 2 } :({\frac {\partial \phi(z)} {\partial z}})^2: \, ,  
\qquad 
{\bar T}({\bar z}) = - {\frac 1 2} :({\bar \partial} {\bar \phi}({\bar z}))^2: 
\ee  
Here $: \, :$ denotes a proper normal ordering. 
Then, through the Laurent expansion of powers of $z$, 
we have 
\be 
L_0 = {\frac 1 2} a_0^2 + \sum_{n=1}^{\infty} a_{-n}a_{n}   
\ee
Thus, the conformal weights $h_{nm}$ and ${\bar h}_{nm}$ are given by 
\be 
h_{n,m} = {\frac 1 2} \left( {\frac n  R} + 
{\frac 1 2} \, m R \right)^2 \, ,  
\qquad 
{\bar h}_{n,m} = {\frac 1 2} \left( {\frac n  R} - 
{\frac 1 2} \, m R \right)^2  
\label{weights}
\ee

\subsubsection{The XXZ spin chain and CFT with $c=1$ }

\par
We  discuss the finite-size corrections to the XXZ spin chain.  
The finite-size corrections to the ground state energy  
is calculated  in \cite{deVega-Karowski,Tsvelik,BIR,deVega-nested} 
 based on the method 
 with the Euler-MacLaurin formula \cite{deVega-Woynarovich}.  
(For a review, see \cite{Korepin,deVega,Nagao}.) 
The result is 
\bea
E_{ex} & = & L e_{\infty} - {\frac {\pi v }{6L}} + 
{\frac {2\pi v} L} \left( {(\Delta D)^2} {\xi^2} 
+ {\frac {(\Delta M)^2} {4 \xi^2}} 
+ N + {\bar N}
\right) \\
P_{ex}-P_0 & = & 2 k_F \Delta D + {\frac {2 \pi } L} 
(\Delta D \Delta M +  N - {\bar N} ) 
\eea
Here the term $e_{\infty}$ denotes  
the ground-state energy per site. The Fermi velocity is 
obtained by the derivative of the dressed energy \cite{Korepin} 
with respect to 
the rapidity at the Fermi level.  

\par 
The central charge $c$ is given by 1. $\Delta D$ and $\Delta M$ 
are integers. $\Delta M$ denotes the change in the number of down spins, 
and $\Delta D$ the number of particles jumping over the Fermi sea 
through the backscattering. 
We note that the difference of the conformal weights (\ref{weights}) 
is given by 
$h_{nm} - {\bar h}_{nm}= nm$.  Thus, $\Delta M$ and $\Delta D$ correspond to 
$n$ and $m$ of the $c=1$ CFT, respectively.  
$N$ and ${\bar N}$ are derived from 
the particle-hole excitations near the Fermi surface.  
The Fermi wavenumber $k_F$ is given by $k_F = \pi M /L$ 
where $M$ is the number of down spins. If the dispersion is linear, 
$k_F$ is consistent with the number of particles $M$.

\par 
 The parameter  $\xi$ is given by the dressed charge, which is defined by 
 an integral equation.  We note that the sum of the conformal 
 weights (\ref{weights}) is given by 
 $h_{n,m} + {\bar h}_{n,m}$ = $n^2/R^2 + m^2 R^2/4$.  
Thus, the dressed charge $\xi$ corresponds to the radius 
$R$ of the $c=1$ CFT, $\xi=R/2$. 
Under zero magnetic field, the dressed charge $\xi$ 
or the radius $R$ is given by 
\cite{Luther} 
\be 
R = \left( {\frac {2\pi}  {\pi - \mu}}  
\right)^{1/2} \qquad (0 \le \mu \le \pi) . 
\ee

%
%

\setcounter{equation}{0} 
\setcounter{figure}{0} 
\renewcommand{\theequation}{3.\arabic{equation}}

\renewcommand{\thefigure}{3.\arabic{figure}}
\setcounter{section}{2}

\section{Various integrable models on two-dimensional lattices}

\subsection{Ising model and  Potts model}

\subsubsection{Ising model}

Let us consider the Ising model defined on a square lattice. 
Each  lattice site has spin variable which takes the two values $\pm 1$. 
We denote by $\sigma_j$ the spin variable of lattice site $j$. 
The Hamiltonian of the Ising model is given by 
\be 
{\cal H} = - J \sum_{\la i,j \ra} \sigma_i \sigma_j 
\ee
where the symbol $\la i,j \ra$ denote that sites $i$ and $j$ 
are  nearest neighbors, and we take the sum over all the 
pairs of adjacent sites on the lattice. 

\par 
There have been many papers written on the two-dimensional 
Ising model \cite{McCoy-Wu}. However, it should be remarked that 
correlation functions are calculated exactly 
 for the two-dimensional Ising model in the scaling limit \cite{Wu,Sato}. 

\par 
We remark that exact solutions are discussed for the Ising model 
defined on various two-dimensional lattices such as the 
Kagome lattice (For a review, see Ref. \cite{Syoji} ).

\subsubsection{Self-dual Potts model}

\par 
The Potts model generalizes the Ising model 
into a $p$-state model with $p > 2$ \cite{Baxter-book,Potts,Kihara}. 
Let us consider the three-state Potts model  defined on a square lattice 
\cite{Potts}. 
Each  lattice site has spin variable which takes three  values 1, 2, 3. 
We denote by $\sigma_j$ the spin variable of lattice site $j$. 
The Hamiltonian of the Potts model is given by 
\be 
{\cal H} = - J \sum_{\la i,j \ra} \delta(\sigma_i, \sigma_j) 
\ee
Here the symbol $\delta(a,b)$ denote the Kronecker delta. 

\par 
In general, the Potts model is not solvable. However, at its criticality, 
it is equivalent to some variant of the six-vertex model, and  is solvable. 
Thus,  the self-dual Potts model is integrable \cite{TL,Inversion}.  

\subsubsection{Ashkin-Teller model}

With each site $i$ we associate two spins: $s_i$ and $t_i$.  
They take values $\pm 1$. The Hamiltonian of the model is given by 
\be
{\cal H} = - \sum_{\la i,j \ra} \left[ K_2 (s_i s_j + t_i t_j) 
+ K_4 s_i s_j t_i t_j \right]
\ee
It is known that 
the Ashkin-Teller model and the eight-vertex model 
are in the same universality class \cite{Kadanoff}. 
The universality class is described by the $c=1$ CFT with 
the twisted boson \cite{SKYang}. 

\subsection{Chiral Potts model}

\subsubsection{General case}
There is another version of the Potts model \cite{Baxter-book,Potts}. 
We may assign the chiral symmetry, or  ${\bf Z}/p{\bf Z}$ symmetry 
on the Potts model, which we  call the chiral Potts model.  

\par 
We now explain the most general chiral Potts model defined on a square lattice
\cite{Perk-review}. Let $a$ and $b$ denote the spin variables 
defined on two nearest-neighboring sites. The interaction energy between the
 spins depends on the difference $n=a-b$ (mod $N$) as 
\be 
{\cal E }(n) = \sum_{j=1}^{N-1} E_j \omega^{jn}   \quad \omega = e^{2 \pi /N}
\ee 
We note that ${\cal E}(N+n) = {\cal E}(n)$. 
The parameter $E_j$ can be written as 
\be 
{\frac {E_j} {k_B T}} =- K_j \omega^{\Delta_j}  \, , \quad {\rm for } \quad j=1, \cdots 
[[{\frac N 2}]] 
\ee
where $K_j$ and $\Delta_j$ constitute $N-1$ independent variables, 
and the symbol $[[\cdot]]$ denotes the Gaussian  symbol.    
For a real number $x$, $[[x]]$ denotes the biggest integer  
 not larger than $x$. 
We also assume that $E_{N-j}$ is complex conjugate to 
$E_j$: $E_{N-j}=E_j^{*}$.  
 When $N$ is odd, we have 
 \be 
- { \frac {{\cal E}(n)} {k_B T}} = 
\sum_{j=1}^{[[(N-1)/2]]} 2 K_j 
\cos \left( {{2\pi} \over N}(jn + \Delta_j) \right) 
+ K_{N/2} (-1)^n \, {\frac {((-1)^N +1)} 2}   
\ee

\subsubsection{Integrable chiral Potts model}

Let us discuss the integrable restriction 
of the most general chiral Potts model 
defined on the square lattice. It has  horizontal and vertical couplings. 
Suppose that spin variables $a$ and $b$ are located on two neighboring sites 
connected by a horizontal line. When the line goes rightward from $a$ to $b$, 
 the horizontal coupling has the energy  ${\cal E}_{pq}(a-b)$ 
 and the Boltzmann weight 
 $W_{pq}(n)$. Here $p$ and $q$ are `rapidity' parameters.  
For spin variables $c$ and $d$  located on two neighboring sites 
connected by a vertical line, the vertical coupling has the 
energy ${\bar {\cal E}}_{pq}(c-d)$ 
and the Boltzmann weight ${\bar W}_{pq}(c-d)$ 
when the vertical line goes upward from $c$ to $d$.  

\par 
The model is called solvable 
if the Boltzmann weights $W$'s and ${\bar W}$'s 
satisfy the star-triangle equation 
\be 
\sum_{d=1}^{N} {\bar W}_{qr}(b-d) W_{pr}(a-d) {\bar W}_{pq}(d-c)
= R_{pqr} W_{pq}(a-b) {\bar W}_{pr})(b-c) W_{qr}(a-c)  
\ee
The solution to the above equation is given by 
\bea 
W_{pq}(n) & = & W_{pq}(0) \prod_{j=1}^{n} 
\left( {\frac {\mu_p} {\mu_q}} \cdot 
{\frac {y_q  - x_p \omega^j}
        {y_p  - x_q \omega^j}} \right) \non \\
{\bar W}_{pq}(n) &  = & {\bar W}_{pq}(0) \prod_{j=1}^{n} 
\left( {\mu_p \mu_q} \cdot {\frac {\omega x_p - x_q \omega^j}
        {y_q - y_p \omega^j}} \right)
\eea
with a constant $R_{pqr}$ depending on the three rapidity variables. 
The constraint that the Boltzmann weight should have  
periodicity modulo $N$: $W_{pq}(n+N)= W_{pq}(n)$ gives 
for all rapidity pairs $p$ and $q$  
\be 
\left( {\frac {\mu_p} {\mu_q}} \right)^N   =  
{\frac {y_p^N - x_q^N} {y_q^N - x_p^N}} \, , \qquad 
\left( {\mu_p} {\mu_q} \right)^N   =  
{\frac {y_q^N - y_p^N} {x_p^N - x_q^N}} \label{cc}
\ee
We can define $k$ and $k^{'}$ such that 
\bea 
\mu_p & = & {\frac {k^{'}} {1 -k x_p^N}} 
= {\frac {1 -k y_p^N}  {k^{'}}} \label{m1} \\
& & x_p^N + y_p^N = k(1+ x_p^N y_p^N) \label{m2}
\eea
where $k^2+ (k^{'})^2 =1$. 
Thus, the  rapidities 
are placed on a curve of genus $g >1$ of Fermat type.  

\par We make some comments. 
Multiplying the  two equations of (\ref{cc}) and noting 
that $p$ and $q$ are independent,  we can show 
eq. (\ref{m1}) and then eq. (\ref{m2}). 
The star-triangle equation 
is proven illustratively in the appendix of Ref. \cite{Kino-Perk}. 
We note that the vertex-type formulation of 
the chiral Potts model is  related to the tetrahedron equation 
\cite{Sergeev}. 
The integrable chiral Potts model generalizes 
the self-dual $Z_N$ model given by Fateev and Zamolodchikov \cite{Fateev}. 
An elliptic extension of the self-dual $Z_N$ model is introduced 
in Ref. \cite{Kashiwara-Miwa}, and the Yang-Baxter equation 
is proven for the model in Ref. \cite{Hasegawa-Yamada}. 
Some nontrivial connections among 
 higher-rank chiral Potts models, elliptic IRF models and 
 Belavin's $Z_N$ symmetric model are explicitly discussed 
 in Ref. \cite{Yamada}.

\subsection{The eight-vertex model}

Let us explain the eight-vertex model which was solved by Baxter 
in 1972 \cite{Baxter72}. 
The Boltzmann weights $w(\alpha, \beta | \gamma, \delta; u)$ 
of  are nonzero if the charge is conserved modulo 2: 
$\alpha + \beta = \gamma + \delta$ (mod) 2. We have  
the six configurations around a vertex shown in Fig. 2.2 and 
the  two in Fig. 3.1 .   We assume that there is no external field. 
The weights therefore become symmetric: $\epsilon_1=\epsilon_2$,  
$\epsilon_3=\epsilon_4$, $\epsilon_5=\epsilon_6$, and $\epsilon_7=\epsilon_8$. 

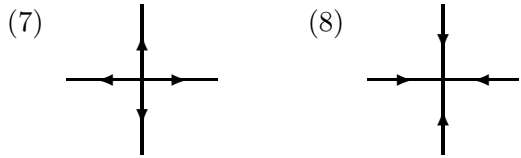
\begin{figure}[htb]
\begin{center}
\setlength{\unitlength}{1mm}
\begin{picture}(120,35)(0,30)
\multiput(30,40)(40,0){2}{{\thicklines\line(-1,0){10}}}
\multiput(40,40)(40,0){2}{{\thicklines\line(-1,0){10}}}
\multiput(30,40)(40,0){2}{{\thicklines\line(0,1){10}}}
\multiput(30,30)(40,0){2}{{\thicklines\line(0,1){10}}}
%
\put(30,40){\thicklines\vector(1,0){6}} 
\put(30,40){\thicklines\vector(-1,0){6}} 
\put(30,40){\thicklines\vector(0,1){6}} 
\put(30,40){\thicklines\vector(0,-1){6}} 
%
\put(60,40){\thicklines\vector(1,0){6}} 
\put(80,40){\thicklines\vector(-1,0){6}} 
\put(70,50){\thicklines\vector(0,-1){6}} 
\put(70,30){\thicklines\vector(0,1){6}} 
\put(12,45){\makebox(5,5){(7)}}
\put(52,45){\makebox(5,5){(8)}}
\end{picture}
\end{center}
\caption{Vertex configurations for the eight-vertex model. (7) $w(2,2|1,1)$; 
(8) $w(1,1| 2,2)$.}
\end{figure} 

\bea 
w(1, 1 | 1, 1) & = & w(2, 2 | 2, 2) = w_1 = a_{8V}   \non \\ 
w(1, 2 | 2, 1) & = & w(2, 1 | 1, 2) = w_2 = b_{8V} \non \\
w(1, 2 | 1, 2) & = & w(2, 1 | 2, 1) = w_3 =c_{8V} \non \\ 
w(1, 1 | 2, 2) & = & w(2, 2 | 1, 1) = w_4 =d_{8V} 
\label{symmetry-8V}
\eea

We give a parametrization of the Boltzmann weights. 
 We define the theta function 
\be 
\theta(z; \tau) = 2 p^{1/4} \sin \pi z \prod_{n=1}^{\infty}
(1-p^{2n})(1-p^{2n}\exp(2\pi i z)) (1-p^{2n}\exp(-2\pi i z)) \, , 
\ee
where the nome $p$ is related to  the parameter $\tau$ by $p=\exp(\pi i \tau)$  
with  ${\rm Im} \quad \tau>0$ . We also define 
the theta functions $\theta_0(z)$ and $\theta_1(z)$ satisfying 
$\theta_{\al}(z+1)= (-1)^{\al} \theta_{\al}(z)$ and 
 $\theta_{\al}(z+\tau)= i e^{-\pi i (z+ \tau/2)} \theta_{1-\al}(z)$ 
 for $\alpha= 0,1$ , 
 where we define $\theta_1(z)$ by $\theta_1(z; \tau)= \theta(z; 2 \tau)$. 
The Boltzmann weights $a_{8V}(z), b_{8V}(z), c_{8V}(z)$ 
and $d_{8V}(z)$ are expressed as   
\bea 
a_{8V}(z)& = & {\frac {\theta_0(z) \theta_0(2\eta)} 
                {\theta_0(z- 2\eta) \theta_0(0)}} \, , \quad 
b_{8V}(z) = {\frac {\theta_1(z) \theta_0(2\eta)} 
                {\theta_1(z- 2\eta) \theta_0(0)}} \, ,  \non \\
c_{8V}(z) & = & - {\frac {\theta_0(z) \theta_1(2\eta)} 
                {\theta_1(z- 2\eta) \theta_0(0)}} \,, \quad 
d_{8V}(z) = - {\frac {\theta_1(z) \theta_1(2\eta)} 
                {\theta_0(z- 2\eta) \theta_0(0)}}  \, . 
\eea

\subsection{IRF models}

\subsubsection{Unrestricted 8V SOS model}

We introduce unrestricted Solid-on-Solid models.  
They are also called Interaction-Round-a-Face models or 
IRF models, briefly \cite{Baxter-book}. 

\par 
To each site $i$ of a two-dimensional square lattice,  
 a spin  $a_i$ is associated.  Let  $i, j, k$, and $\ell$ 
 be the lattice sites surrounding a face (or a square),  
 where $i,j,k, \ell$ are placed counterclockwise from  the southwest corner. 
    We assume that an elementary configuration  is given by 
that of the four spin variables  around the face, and the probability 
of having $a_i, a_j, a_k, a_{\ell}$ 
 is  denoted  by the Boltzmann weight 
$w(a_i, a_j, a_k, a_{\ell}; z)$.  
Here the variable $z$ is called the spectral parameter.

\par 
 For unrestricted 8V SOS model,  $a_i$ can take any integer.  
When sites $i$ and $j$ are   nearest neighboring, then 
the states $a_i$ and $a_j$ are said to be admissible if and only if 
$|a_i - a_j| = 1$ \cite{Baxter73,Baxter-book}.    
The Yang-Baxter equations are given by   
\bea
& & \sum_g w(a,b,g,f; z-w) w(f,g,d,e; z) w(g,b,c,d; w) \non \\ 
& = & \sum_g w(f,a,g,e; w) w(a,b,c,g; z) w(g,c,d,e; z-w) \, , 
\label{ybrIRF}
\eea
where the summation of the variable $g$ is taken over all 
the admissible states.  
\par 
The Boltzmann weights are given by 
\bea 
w(d+1,d+2,d+1,d; z, w_0 ) & = &  w(d,d-1,d-2,d-1; z ) = 
{\frac {\theta(2\eta -z)} {\theta(2\eta)}} 
  \non \\ 
w(d-1,d,d+1,d; z, w_0 ) & = &   w(d+1,d,d-1,d; z, w_0 ) \non \\
& = & {\frac {\theta(z)} {\theta(2\eta)}} \, 
{\frac {\sqrt{\theta(2\eta (d+1) + w_0 ) \theta(2\eta (d-1) + w_0)}} 
{\theta(2\eta d + w_0)} } \non \\ 
w(d+1,d,d+1,d; z, w_0 ) & = &  
{\frac {\theta(z+ 2\eta d + w_0)}  {\theta(2\eta d + w_0)}}    \non \\ 
 w(d-1,d,d-1,d; z,w_0 ) & = &   
 {\frac {\theta(z- 2\eta d - w_0)}  {\theta(2\eta d + w_0)}}    
\label{YBE-IRF}
\eea

\subsubsection{RSOS models}

Let us explain restricted Solid-on-Solid models (RSOS models) \cite{ABF}.  
Let $s$ denote the number of elements in $S$. Consider a $s \times s$ matrix $C$ 
satisfying the following conditions \cite{Kuniba,Akutsu}: 

\par 
(i) $C_{ab} = C_{ba} = 0$ \, {\rm or} \,  1 
\par 
(ii) $C_{aa} =0$ 
\par 
(iii) For each $a \in S$, there should exist $b \in S$ such that $C_{ab} =1$ 

\par \noindent 
 For such choice of $C$, we impose a restriction 
that two states $a$ and $b$ can occupy the neighboring lattice sites 
if and only if $C_{ab}=1$. We call such a pair 
of the states $(a,b)$ admissible. 
 For the case of unrestricted models, 
 the infinite matrix $C$ satisfies the conditions (i), (ii), 
 and (iii) with an infinite set $S$. 

\par 
 For an illustration, let us consider 
 the restricted eight-vertex Solid-on-Solid model 
(the restricted 8V SOS model), which we also call the ABF model \cite{ABF}.  
 For the $N$-state case, we have $S=\{1, 2, \ldots, N\}$. 
The nonzero matrix elements of $C$ are 
given by $C_{j,j+1}=C_{j+1,j}=1$ for $j=1,2, \ldots, N-1$;   
other matrix elements such as $C_{1, N}$ and $C_{N,1}$  are 
given by zero.  
Setting $w_0=0$, 
then we have the Boltzmann weights of the ABF model. 
The Boltzmann weights 
satisfy the Yang-Baxter relations (\ref{ybrIRF}) 
with the finite set: $S= \{1,2, \ldots, N \}$.

\par 
Let us explain CSOS models,  another type of RSOS models \cite{Pearce,Kuniba,Akutsu}.  
Here we assume that $2 N \eta = m_1$, where integer $m_1$ has no common 
divisor with  $N$. 
If we set $w_0 \neq 0$, 
then we have the Boltzmann weights of the cyclic SOS model (CSOS model). 
We can show that the Boltzmann weights  
satisfy the Yang-Baxter relations 
with the finite set: $S= \{1,2, \ldots, N \}$ and the 
cyclic admissible conditions: $C_{1,N} =C_{N,1} = 1$. 

\par 
We remark that the connection between the 6j symbols and the Boltzmann 
weights of IRF models is first discussed in Ref. \cite{Pasquier}.

\subsubsection{Fusion IRF models and ABCD IRF models}

The 8VSOS model was generalized into the fusion IRF models. 
\cite{Fusion-IRF} IRF models associated with $A_n^{(1)}$ Lie algebra 
are constructed. 
\cite{A-IRF} The 
IRF models associated with $ B^{(1)}$ $C^{(1)}$ $D^{(1)}$ typed Lie algebra 
are also obtained \cite{ABCD}. 

\par 
In the IRF models, the one-point function, which is the magnetization per site, 
can be calculated by the corner transfer matrix method 
invented by Baxter \cite{Baxter-book}.

\subsubsection{Gauge transformations }
\par 
It is sometimes convenient to employ a gauge transformation   
\be 
w(a,b,c,d ; z) \rightarrow w(a,b,c,d; z) \, {\frac {g_c} {g_a} } \, . 
\label{gauge-t}
\ee
The transformed Boltzmann weights also satisfy 
the Yang-Baxter relations (\ref{ybrIRF}). For instance, we may 
set $g_a = \exp(\pi i a/2) \, \sqrt{\theta(2 \eta a + w_0) }$ 
($a \in {\bf Z}$).

%
%

\setcounter{equation}{0} 
\setcounter{figure}{0} 
\renewcommand{\theequation}{4.\arabic{equation}}
\renewcommand{\thefigure}{4.\arabic{figure}}
\setcounter{section}{3}

\section{Yang-Baxter equation and the algebraic Bethe Ansatz }

\subsection{Solutions to the Yang-Baxter equation} 

\subsubsection{Derivation of a solution for the six-vertex model}

\par 
Let us solve the Yang-Baxter equation for the six-vertex model. 
We recall that it is given by 
\bea 
& & \sum_{\al,\bt,\gm} w(\al, \gm | a_1, a_2)
w^{'}(\bt, b_3 | \gm, a_3) w^{''}(b_1, b_2 | \al, \bt ) \non \\
& = & \sum_{\al, \bt, \gm} w^{''}(\bt, \al | a_2, a_3 ) 
w^{'}(b_1, \gm | a_1, \bt) w(b_2, b_3 | \gm, \al) \, . \label{ybr} 
\eea
The Yang-Baxter equation is illustrated in Fig. 4.1.

\par 
There are $2^3 \times 2^3 = 64$ cases 
 for the entries $(a_1, a_2, a_3)$  and  $(b_1, b_2, b_3)$.  
Due to the ice rule, however, the Yang-Baxter equation is trivial 
unless $a_1+ a_2+ a_3= b_1+ b_2 + b_3$. 
We have thus only 20 entries:  
$\left(
 _3C_0 \right)^2 + \left( _3C_1 \right)^2 + \left(
 _3C_2 \right)^2 + \left( _3C_3 \right)^2  
= 1 + 9 + 9 + 1 = 20$.

%
%
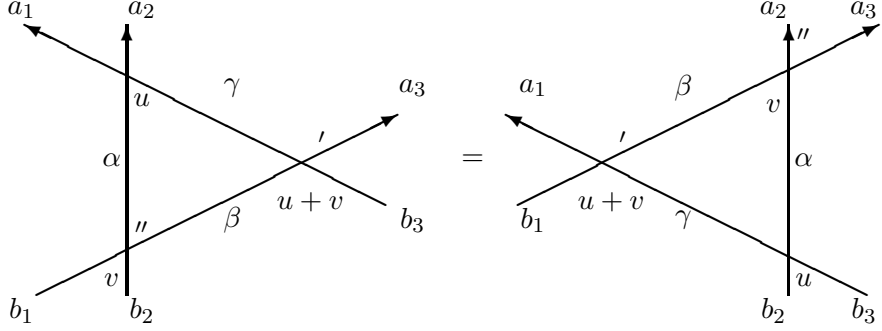
\begin{figure}[htb]     
\begin{center}
\setlength{\unitlength}{0.8mm}
\begin{picture}(160,55)(0,0)
\put(10,5){\thicklines\vector(2,1){60}}
\put(68,20){\thicklines\vector(-2,1){60}}
\put(25,5){\thicklines\vector(0,1){45}}
\put(90,20){\thicklines\vector(2,1){60}}
\put(148,5){\thicklines\vector(-2,1){60}}
\put(135,5){\thicklines\vector(0,1){45}}
\put(5,50){\makebox(5,5){$a_1$}}
\put(25,50){\makebox(5,5){$a_2$}}
\put(70,37){\makebox(5,5){$a_3$}}
\put(80,25){\makebox(5,5){$=$}} 
\put(90,37){\makebox(5,5){$a_1$}}
\put(130,50){\makebox(5,5){$a_2$}}
\put(145,50){\makebox(5,5){$a_3$}}
\put(5,0){\makebox(5,5){$b_1$}}
\put(25,0){\makebox(5,5){$b_2$}}
\put(70,15){\makebox(5,5){$b_3$}}
\put(90,15){\makebox(5,5){$b_1$}}
\put(130,0){\makebox(5,5){$b_2$}}
\put(145,0){\makebox(5,5){$b_3$}}
\put(20,25){\makebox(5,5){$\alpha$}}
\put(40,37){\makebox(5,5){$\gamma$}}
\put(40,15){\makebox(5,5){$\beta$}}
\put(135,25){\makebox(5,5){$\alpha$}}
\put(115,37){\makebox(5,5){$\beta$}}
\put(115,15){\makebox(5,5){$\gamma$}}
\put(55,28){\makebox(5,5){$'$}}
\put(25,13){\makebox(5,5){$''$}}
\put(105,28){\makebox(5,5){$'$}}
\put(135,45){\makebox(5,5){$''$}}
\put(25,35){\makebox(5,5){$u$}}
\put(53,18){\makebox(5,5){$u+v$}}
\put(20,5){\makebox(5,5){$v$}}
\put(135,5){\makebox(5,5){$u$}}
\put(103,18){\makebox(5,5){$u+v$}}
\put(130,34){\makebox(5,5){$v$}}
%
%
%
%
%
%
\end{picture}
\end{center}
\caption{The Yang-Baxter equations for  vertex models. 
The spectral parameters are shown 
by the angles between pairs of straight lines. } 
\end{figure}

\par 
Let us consider  symmetries of the Boltzmann weights 
given in (\ref{symmetry}). 
If we exchange $1$ and $2$,  
the Boltzmann weights do not change.  
Thus, we have reduced 20 cases into 10 cases. 
\bea 
& & (a_1,a_2,a_3; b_1,b_2,b_3)  =  (1,1,1; 1,1,1), \non \\
& & \qquad  (1,1,2;1,1,2), (1,1,2; 1,2,1), (1,1,2; 2,1,1), \non \\
& & \qquad  (1,2,1; 1,1,2), (1,2,1; 1,2,1), (1,2,1; 2,1,1), \non \\
& & \qquad  (2,1,1; 1,1,2), (2,1,1; 1,2,1) , (2,1,1; 2,1,1) 
\eea
We  now recall that the Boltzmann weights (\ref{symmetry}) have 
 the symmetry  
\be
w(\al,\bt | \gm, \dt) = w(\gm, \dt | \al,\bt ) = w(\bt,\al | \dt, \gm)
\label{sym}
\ee
Combining these symmetries we can show that the Yang-Baxter equation for the 
two entries (1) and (2) are equivalent: 
(1) $(a_1,  a_2,  a_3;  b_1,  b_2,  b_3 )$ ; 
(2) $(b_3, b_2,  b_1 ; a_3,  a_2 , a_1 )$. Precisely, 
the L.H.S. (or R.H.S) 
of the Yang-Baxter equation of the case $(1)$ corresponds to 
the R.H.S. (or L.H.S) of eq. (\ref{ybr}). 
Thus, we have the three cases as follows.  
\be 
(1,1,2; 1,1,2), (1,1,2; 1,2,1), (1,2,1; 1,1,2) 
\ee
For the three cases, the Yang-Baxter equations are given by 
\bea
ac^{'}a^{''} & = & bc^{'}b^{''}  +ca^{'}c^{''}  \non \\
ab^{'}c^{''}  & = & ba^{'}c^{''} + cc^{'}b^{''}  \non \\
cb^{'}a^{''}  & = & ca^{'}b^{''}  + bc^{'}c^{''} 
\eea 
A nontrivial solution 
$(a^{''},b^{''},c^{''})$ exists only if the determinant vanishes 
\be \left| 
\begin{array}{ccc}  
a c^{'} & - b c^{'} & - c a^{'} \\
0 & c c^{'} & b a^{'} - a b^{'} \\
c b^{'} & - c a^{'} & -b c^{'} 
\ear
\right| = a b c a^{'} b^{'} c^{'} \, \left( 
{\frac {(a^{'})^2 + (b^{'})^2 -(c^{'})^2} {a^{'} b^{'}} }- 
{\frac {a^2 + b^2 -c^2} {a b } }
\right) 
\ee 
We define the parameter $\Delta$ by   
\be 
\Delta = {\frac {a^2 + b^2 -c^2} {2 a b } }
\ee
The condition that the determinant vanishes is given by 
\be 
\Delta = \Delta^{'} 
\ee
Thus, the transfer matrices $\tau$ and $\tau^{'}$ commute,   
if the two sets of weights $(a,b,c)$ and 
$(a^{'},b^{'},c^{'})$ have the same $\Delta$.

\par 
In terms of the spectral parameter $u$, 
we can parametrize the three Boltzmann weights 
$a$, $b$ and $c$. We express the weight as 
$w(\al, \bt | \gm, \dt; u) $. 
Let $u$ and $v$ be arbitrary. We denote 
$w(\al, \bt | \gm, \dt)$, $w^{'}(\al, \bt | \gm, \dt)$ , and 
$w^{''}(\al, \bt | \gm, \dt)$ as 
$w(\al, \bt | \gm, \dt; u)$, 
$w(\al, \bt | \gm, \dt; u+v)$, and 
$w(\al, \bt | \gm, \dt; v)$, respectively.  
The Yang-Baxter equations are depicted in Fig. 4.2. 
%
As a solution, we may have 
$(a, b, c)  = 
( \rho \sinh(u + 2 \eta),  
  =  \rho \sinh u , 
 =  \rho \sinh 2\eta $ . Here we set 
 $\Delta= \cosh 2\eta $ . 
 The transfer matrices $\tau(u)$ and $\tau(v)$ 
 commute: $\tau(u) \tau(v) = \tau(v) \tau(u)$.

\subsubsection{Gauge transformations for vertex models} 

Let us suppose that $w$'s satisfy the Yang-Baxter equation. 
Then we can show that transformed weights 
${\tilde w}$'s defined by 
\be 
{\tilde w}(\alpha, \beta | \gamma, \delta; u) =    
(\epsilon)^{\alpha + \gamma} \exp(\kappa (\alpha + \gamma - \beta -\delta) u) 
w(\alpha, \beta | \gamma, \delta; u)   
\label{gt-vertex}
\ee
also satisfy the Yang-Baxter equations \cite{AW,WDA}.   
Here $\epsilon= \pm 1$, and the number $\kappa$ is arbitrary. 
\par 
The gauge transformation is important in the derivation of the 
Jones polynomial from the symmetric Boltzmann weights 
of the six-vertex model under zero field \cite{AW}. 
It is also quite useful 
when we discuss the relation of the six-vertex model 
to the quantum group, as we shall see in \S 5 (see also 
the appendix of Ref. \cite{loop}).

\subsection{Algebraic Bethe ansatz}

\subsubsection{$R$ matrix and the $L$ operator}

Let us diagonalize the transfer matrix by the method of 
the algebraic Bethe ansatz.  
\cite{Faddeev-Takhtajan,Takhtajan,Korepin}

\par 
Let us  introduce the notation of the matrix tensor product. 
We  define the direct product $A \otimes B$ of matrices 
$A$ and $B$.  
Let $A^{j}_{k}$ denote the matrix element for the entry of 
column $j$ and row $k$  of the matrix $A$.  
Then, the matrix element of column 
$(j_1, j_2)$ and $(k_1, k_2)$ is defined  by   
\be \left( 
A \otimes B \right)^{j_1, j_2}_{k_1, k_2} = 
A^{j_1}_{k_1} B^{j_2}_{k_2}   
\ee

\par 
We now define the R matrix of the XXZ spin chain. 
The element $R^{ab}_{cd}$ corresponds to the entry of 
column $(a,b)$ and row $(c,d)$.  
\be
R(z)  = \left(
 \begin{array}{cccc} 
 R(z)^{11}_{11} & R(z)^{11}_{12} & R(z)^{11}_{21} &  R(z)^{11}_{22} \\
R(z)^{12}_{11} & R(z)^{12}_{12} & R(z)^{12}_{21} & R(z)^{12}_{22} \\
R(z)^{21}_{11} & R(z)^{21}_{12} & R(z)^{21}_{21} & R(z)^{21}_{22} \\
R(z)^{22}_{11} & R(z)^{22}_{12} & R(z)^{22}_{21} & R(z)^{22}_{22}
\ear
\right) 
= \left(
\begin{array}{cccc} 
a(z) &   0 & 0 & 0 \\
0 &   c(z) & b(z) & 0 \\
0 &   b(z) & c(z) & 0 \\
0 &   0 & 0 & a(z)
\ear
\right) 
\label{Rmatrix}
\ee
Here  $a(z)$, $b(z)$ and $c(z)$ are given by 
\be 
a(z) = \sinh (z + 2 \eta) \, ,  \quad 
b(z)=  \sinh z \, , \quad c(z)= \sinh 2 \eta \, . 
\label{abcRmatrix}
\ee
Here, the functions $a(z)$, $b(z)$ and $c(z)$ are equivalent to 
the Boltzmann weights $a$, $b$ and $c$ in \S 2 

\par 
We now introduce $L$ operators. 
We write the matrix element  the $L$ operator with entry $(j,k)$ 
as  $\left( L_n(z) \right)_{jk}$ or  $L_n(z)^j_k$ . The $L$ operator 
for the XXZ spin chain is given by 
\be 
 L_n(z) 
= \left(
\begin{array}{cc}  
L_n(z)^1_1  &  L_n(z)^1_2 \\
 L_n(z)^2_1  & L_n(z)^2_2  
\ear
\right)  
=  \left(
\begin{array}{cc}  
\sinh \left( z I_n + \eta \sigma_n^z \right) & \sinh 2 \eta \, \sigma_n^{-} \\
\sinh 2 \eta \, \sigma_n^{+}  & \sinh \left( z I_n - \eta \sigma_n^z \right) 
\ear
\right)  
\ee
Here $I_n$ and $\sigma_n^a$ ($n=1, \ldots, L$) are 
acting on the $n$ th vector space $V_n$. 
The $L$ operator is an operator-valued matrix 
which acts on the auxiliary vector space 
$V_0$.   
The symbols $\sigma^{\pm}$ denote 
$\sigma^{+}= E_{12}$ and $\sigma^{-} = E_{21}$,    
and $\sigma^x, \sigma^y, \sigma^z$ are the Pauli matrices  
%
%

\par 
In terms of the $R$ matrix and $L$ operators, 
the Yang-Baxter equation is expressed as 
\be 
R(z-t) \left( L_n(z) \otimes L_n(t) \right) 
 = \left( L_n(t) \otimes L_n(z) \right) R(z-t) 
\label{RLL} 
\ee
Here the tensor symbol in $L_n(z) \otimes L_n(t)$ 
denotes the tensor product of auxiliary spaces.

%
%
\begin{figure}[htb]
\begin{center}
\setlength{\unitlength}{1mm}
\begin{picture}(160,40)(0,-5)
\put(15,25){\makebox(5,5){$(1)$}}  
\put(40,15){\thicklines\vector(1,0){20}}
\put(40,15){\thicklines\line(-1,0){20}}
\put(40,15){\thicklines\vector(0,1){20}}
\put(40,15){\thicklines\line(0,-1){20}}
\put(41,-5){\thicklines\vector(0,1){40}}
\put(15,15){\makebox(5,5){$b$}}
\put(42,-5){\makebox(5,5){$\beta_n$}}
\put(42,30){\makebox(5,5){$\alpha_n$}}
\put(60,15){\makebox(5,5){$a$}}
\put(46,16){\makebox(5,5){$z-\eta$}} 
%
%
\put(70,25){\makebox(5,5){$(2)$}}  
\put(90,0){\thicklines\vector(1,1){30}}
\put(120,0){\thicklines\vector(-1,1){30}}
\put(85,25){\makebox(5,5){$a_1$}}
\put(120,25){\makebox(5,5){$a_2$}}
\put(85,0){\makebox(5,5){$b_1$}}
\put(120,0){\makebox(5,5){$b_2$}}
\put(103,17){\makebox(5,5){$z$}} 
\end{picture}
\end{center}
\caption{(1) $L_n(z)^{a}_{b} |_{\alpha_n, \beta_n}$ ; 
(2) $ R(z)^{a_1, a_2}_{b_1, b_2}$ }
\end{figure}
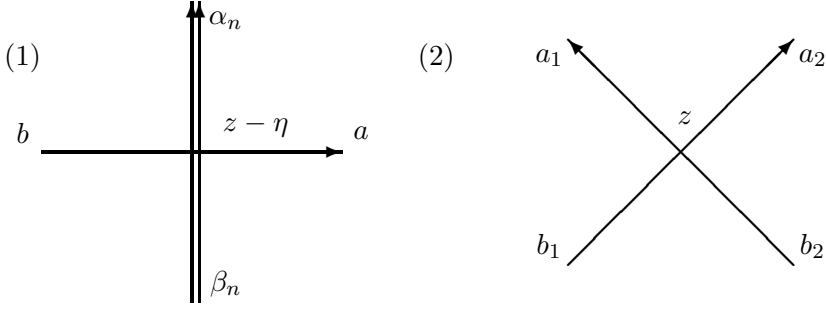

\par 
The Yang-Baxter equation (\ref{RLL}) gives the relation 
between the two products of 
$ 4 \times 4$ matrices.   
For an illustration, we consider the L.H.S. of (\ref{RLL}).  
\bea 
\Big[ R(z-t) \cdot L_n(z) 
\otimes L_n(t) \Big]^{a_1,a_2}_{b_1,b_2}  
& = & \sum_{c_1,c_2} R(z-t)^{a_1, a_2}_{c_1, c_2}
 \, \left(  L_n(z) \otimes 
 L_n(t) \right)^{c_1, c_2}_{b_1, b_2}  
\non \\
& = & \sum_{c_1,c_2} R(z-t)^{a_1, a_2}_{c_1, c_2} \, 
 L_n(z)^{c_1}_{b_1} \times  L_n(t)^{c_2}_{b_2}  
\eea
Here the symbol $\times$ denotes the product of matrices acting on 
the $n$ th space $V_n$. 
%
%
%
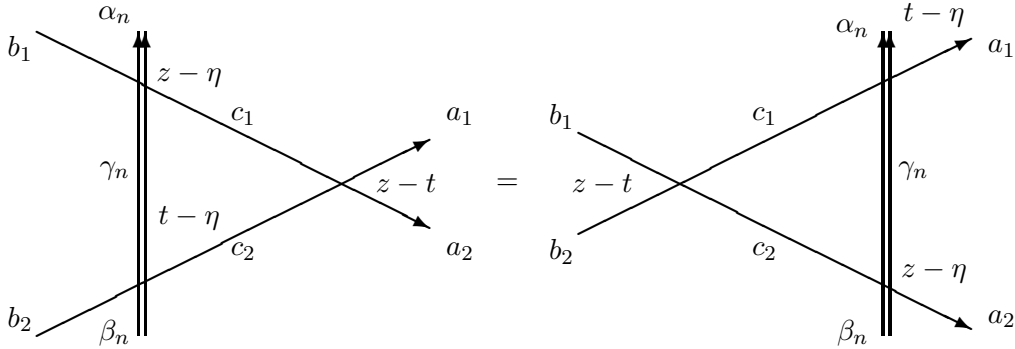
\begin{figure}[htb]
\begin{center}
\setlength{\unitlength}{0.9mm}
\begin{picture}(160,50)(0,5)
%
%
%
%
\put(90,20){\thicklines\vector(2,1){58}}
\put(90,35){\thicklines\vector(2,-1){58}}
\put(135,5){\thicklines\vector(0,1){45}}
\put(136,5){\thicklines\vector(0,1){45}}
\put(91,25){\makebox(5,5){$z-t$}}
\put(140,50){\makebox(5,5){$t-\eta$}}
\put(140,12){\makebox(5,5){$z-\eta$}}
%
%
\put(77,25){\makebox(5,5){$=$}} 
%
\put(10,5){\thicklines\vector(2,1){58}}
\put(10,50){\thicklines\vector(2,-1){58}}
\put(25,5){\thicklines\vector(0,1){45}}
\put(26,5){\thicklines\vector(0,1){45}}
\put(62,25){\makebox(5,5){$z-t$}}
\put(30,20){\makebox(5,5){$t-\eta$}}
\put(30,41){\makebox(5,5){$z-\eta$}}
%
%
%
%
%
%
%
\put(128,48){\makebox(5,5){$\alpha_n$}}
\put(128,3){\makebox(5,5){$\beta_{n}$}}
\put(137,27){\makebox(5,5){$\gamma_n$}}
\put(115,35){\makebox(5,5){$c_1$}}
\put(115,15){\makebox(5,5){$c_2$}}
\put(85,35){\makebox(5,5){$b_1$}}
\put(85,15){\makebox(5,5){$b_2$}}
\put(150,45){\makebox(5,5){$a_1$}}
\put(150,5){\makebox(5,5){$a_2$}}
%
%
%
\put(19,50){\makebox(5,5){$\alpha_n$}}
\put(19,3){\makebox(5,5){$\beta_{n}$}}
\put(19,27){\makebox(5,5){$\gamma_n$}}
\put(38,35){\makebox(5,5){$c_1$}}
\put(38,15){\makebox(5,5){$c_2$}}
\put(5,45){\makebox(5,5){$b_1$}}
\put(5,5){\makebox(5,5){$b_2$}}
\put(70,35){\makebox(5,5){$a_1$}}
\put(70,15){\makebox(5,5){$a_2$}}
\end{picture}
\end{center}
\caption{The Yang-Baxter equation: $R(z-t) (L(z) \otimes L(t)) 
= (L(t) \otimes L(z)) R(z-t)$, eq. (\ref{RLL}). } 
\end{figure} 
%
Expressing  the operator products in the $n$th space $V_n$, 
the L.H.S. of (\ref{RLL}) is written as follows  
\bea 
\Big[ R(z-t) \cdot L_n(z) \otimes L_n(t) 
\Big]^{a_1,a_2}_{b_1,b_2} |_{\alpha_n, \beta_n}   
& = & \sum_{c_1,c_2} R(z-t)^{a_1, a_2}_{c_1, c_2}
 \, \left( 
 L_n(z) \otimes L_n(t) \right)^{c_1, c_2}_{b_1, b_2}  
 |_{\alpha_n, \beta_n} 
\non \\
& = & \sum_{c_1,c_2} \sum_{\gamma_n} 
R(z-t)^{a_1, a_2}_{c_1, c_2} \, 
 L_n(z)^{c_1}_{b_1} |_{\alpha_n, \gamma_n}  \, 
 L_n(t)^{c_2}_{b_2} |_{\gamma_n, \beta_n} 
\eea

\subsubsection{Monodromy matrix and the construction of the eigenvector } 

We define the monodromy matrix by the product of the $L$ operators 
(see also Fig. A.1)
\be
T(z) = L_N(z) \cdots L_2(z) L_1(z) \, . 
\ee
The transfer matrix $\tau_{6V}(z)$ 
of the six vertex model is given by the trace of $T(z)$ 
\be
\tau_{6V}(z) = {\rm tr} T(z) = A(z) + D(z) \, , \quad {\rm where} \quad  
T(z) = \left( 
\begin{array}{cc}  
A(z) & B(z) \\ 
C(z) & D(z) 
\ear
\right)
\ee
The Yang-Baxter equation (\ref{RLL}) leads to 
the commutation relation: $R(z-t) (T(z) \otimes T(t)) = 
(T(t) \otimes T(z)) R(z-t)$,  from which we have many 
 relations among the operators $A$, $B$, $C$, and $D$. 
For instance, we have $B(z)B(t)= B(t)B(z)$. Furthermore, we have  
\bea 
A(z) B(t) & = & 
{\frac {a(t-z)}{b(t-z)} } B(t)A(z) 
- {\frac  {c(t-z)} {b(t-z)}} B(z) A(t) 
\label{AB} \\
D(z) B(t) & = & 
{\frac  {a(z-t)} {b(z-t)} } B(t) D(z) 
- {\frac  {c(z-t)} {b(z-t)}} B(z) D(t) 
\label{DB}
\eea
We define the `vacuum' by 
\be 
|0 \rangle = \overbrace{ | \uparrow \ra_1  | \uparrow \ra_2  
 \cdots  | \uparrow \rangle_N }^{ N}  
\ee
Multiplying $A(z)$ and $D(z)$ on the vacuum,  we have 
\be 
A(z) | 0 \ra = a(z- \eta)^N | 0 \ra \, , \quad 
D(z) | 0 \ra = b(z- \eta)^N | 0 \ra \, . 
\ee

\par 
Let us consider the vector generated by the product of $B$ operators. 
\be 
| M \ra = B(t_1) \cdots B(t_M) | 0 \ra 
\ee
Then, through the commutation relations such as (\ref{AB}) and (\ref{DB}) 
we can show \cite{8VABA,Korepin} that the vector $|M \ra$ gives an 
eigenvector of the transfer matrix $\tau_{6V}(z)$ 
if rapidities $t_1, t_2,\ldots$, 
$t_M$ satisfy the set of equations   
\be 
\left( {\frac {a(t_j-\eta)}  {b(t_j- \eta)}} \right)^N 
= - {\frac {c(t_k-t_j)} {b(t_k-t_j)}} {\frac {b(t_j-t_k)} {c(t_j-t_k)}} 
   \prod_{k=1; \, k \ne j}^{M} \left( {\frac {a(t_j- t_k)} {b(t_j-t_k)}} 
    {\frac {b(t_k- t_j)} {a(t_k-t_j)}} \right) \, , \quad  
   {\mbox{ for}} \quad  j=1, \ldots, M. 
  \label{preBAE} 
%
\ee
They are the Bethe ansatz equations (\ref{BAE}) 
with different parametrization.   
For a set of solutions,  $t_1, t_2, \ldots, t_M$ to eqs. (\ref{preBAE}),  
the eigenvalue of the transfer matrix $\tau_{6V}(z)$ is given by 
\bea 
\Lambda(z; t_1, t_2, \ldots, t_M) & = & 
  a(z - \eta)^N \prod_{j=1}^{M} {\frac {a(t_j -z)} {b(t_j-z)}} +  
 b(z - \eta)^N \prod_{j=1}^{M} {\frac {a(z-t_j)} {b(z-t_j)}} 
 \non \\
& = & \sinh^N(z + \eta) \prod_{j=1}^{M} 
{\frac { \sinh(t_j-z + 2 \eta) } {\sinh(t_j-z) }}
+ \sinh^N(z - \eta) \prod_{j=1}^{M} 
{\frac {\sinh(t_j- z - 2 \eta)} {\sinh(t_j - z ) }} \, . \non \\
\label{evABA}
\eea

\subsubsection{Connection to the coordinate Bethe ansatz result}
\par 
Let us compare the result of the coordinate Bethe ansatz in \S 2. 
We consider the disordered phase: $-1 < \Delta < 1$.  
First, we change the variables $w$ and $\al$ defined in \S 2 
into $u$ and $\zeta$ by 
$w = 2 u - \mu$ and $\alpha = 2 \zeta - i \mu$, respectively. 
Thus, from the expressions (\ref{weightAF}) we have 
$(a, b,c)$ = $(\sin(\mu -u), \sin u, \sin \mu)$ = 
$i(\sinh(-i\mu + i u), \sinh(-iu), \sinh(-i\mu) )$ and    
\be
 L^{Baxter}(z_j) = {\frac {ab + (c^2-b^2)z_j} {a(a-bz_j)}}
 = - {\frac {\sinh((\alpha_j-i w - 2i \mu)/2)} {\sinh((\alpha_j - iw)/2)}}  
=  {\frac {\sinh(-(\zeta_j-i u)  + i \mu)} {\sinh(\zeta_j - iu)}}  
\label{Lbax}
 \ee 
Here we recall that 
the symbol 
$L^{Baxter}(z)$ has been defined in eq. (\ref{LM}) 
in order to denote the eigenvalue of the transfer matrix 
of the six-vertex model.   
The expression (\ref{Lbax}) can be derived from the formula (\ref{evABA}) 
of the algebraic Bethe ansatz as follows.  First,  we take  
 the gauge transformation: $b(z) \rightarrow - b(z)$, 
 which corresponds to the case $\epsilon=-1$ and $\kappa =0$ 
 in eq. (\ref{gt-vertex}).  
Then, we replace the variables $z$, $t_j$'s, and $2 \eta$ in (\ref{evABA})  
by $z-\eta \rightarrow -iu$,  
$t_j-\eta \rightarrow - \zeta_j$, 
and $2 \eta \rightarrow i \mu$, respectively.   
We  have the following:  
\be 
{\frac {a(t_j -z)} {b(t_j-z)}} 
\rightarrow 
- {\frac {a(t_j -z)} {b(t_j-z)}} 
= - {\frac { \sinh(t_j -z  + 2 \eta) } {\sinh(t_j-z) }}
\rightarrow 
 {\frac { \sinh(-(\zeta_j - iu)  + i \mu) } {\sinh(\zeta_j-iu) }}
\ee
and $\sinh^N(z + \eta) \rightarrow \sinh^N(-iu + i \mu)$ . 
Thus, the formula (\ref{evABA}) reproduces the 
expression (\ref{evBA}) for the eigenvalues 
of the transfer matrix except for 
the normalization factor $i^N$.

%
%


\setcounter{equation}{0} 
\renewcommand{\theequation}{5.\arabic{equation}}
\setcounter{section}{4}

\section{Mathematical structures of integrable lattice models}

\subsection{Braid group}

\subsubsection{The Yang-Baxter equation in operator formalism}

\par 
The Yang-Baxter equation in \S 2 gives a  
sufficient condition for the existence of 
 commuting transfer matrices. 
However, there are other viewpoints on 
the Yang-Baxter equation. 

\par 
Let $E^{j}_k$ denote the matrix given by 
\be \left( 
 E^j_k \right)^{a}_{b} = \delta_{a, j} \delta_{b, k} 
\quad a, b = 1, 2 \, . 
\ee
We define operators $X_j(u)$'s by 
\be 
X_j(z) = \sum_{a,b,c,d} w(a,b | c,d; z) 
\overbrace{I \otimes \cdots \otimes I}^{\otimes (j-1)} 
 \otimes E^{c}_{a} \otimes E^{d}_{b}
\otimes \overbrace{I \otimes \cdots \otimes I}^{\otimes (N-j-1)}  
 \quad {\mbox{ for}} \quad j = 1, \ldots, N-1. 
\ee
Here, the symbol $w(a, b | c, d ; z)$ corresponds to $R_{a,b}^{c,d}(z)$, 
and the essential part of $X_j(z)$ is given by 
\be 
X(z) = \sum_{a,b,c,d} w(a,b | c,d; z) E^{c}_{a} \otimes E^{d}_{b} 
\ee
which is equivalent to $R(z)$.  
In terms of the $X_j(z)$'s,  the  Yang-Baxter equation 
is expressed by the following  
\be
X_{j}(z) X_{j+1}(z+t) X_j(t) = 
 X_{j+1}(t) X_j(z+t) X_{j+1}(z) \, , \quad {\rm for} \quad 
j=1, \ldots, N-1 \, . 
\label{operatorYB}
\ee

\subsubsection{The braid group}

\par 
The braid group $B_N$ for $N$ strings is an infinite group 
which is generated 
by the generators $b_1, \ldots, b_{N-1}$ satisfying 
the defining relations 
\bea 
b_j b_{j + 1 } b_j & = & b_{j+1} b_j b_{j+1} \non \\ 
b_i b_j & = & b_j b_i \quad {\rm for} \quad |i-j| > 1  
\eea

\par 
Let us assume that  the limit: $\lim_{z \rightarrow \infty}X_j(z)$ 
exists. Then, eq. (\ref{operatorYB}) becomes
\be 
X_{j}(\infty) X_{j+1}(\infty) X_j(\infty) = 
 X_{j+1}(\infty) X_j(\infty) X_{j+1}(\infty) \, , \quad {\rm for} \quad 
j=1, \ldots, N-1 \, . 
\ee
This is nothing but the defining relations of the braid group. 
Thus, the Boltzmann weights of solvable models 
expressed in terms of the spectral parameter 
lead to representations of the braid group. 

\par 
We now show that from a given exactly solvable model, one can 
derive two different representations of the braid group \cite{AW}. 
Here, the gauge transformation (\ref{gt-vertex}) plyas a central role. 
This technical point is quite fundamental when we make connections 
of exactly solvable models to quantum groups 
and the Temperley-Lieb algebra (for instance, 
see the appendix of Ref. \cite{loop}). 
\par 
We first consider the Boltzmann weights of the zero-field six-vertex model 
given by eqs. (\ref{abcRmatrix}).   
Taking  the infinite limit to them, 
we have the following 
\be
\lim_{z \rightarrow \infty} X(z)/\sinh(z + 2 \eta) = \left(
\begin{array}{cccc} 
1 &   0 & 0 & 0 \\
0 &  0 & \exp(- 2 \eta) & 0 \\
0 &   \exp(- 2 \eta) & 0 & 0 \\
0 &   0 &  0  &  1
\ear
\right) 
\label{bm1}
\ee
The representation of the braid group (\ref{bm1}) leads to a  
 link polynomial equivalent to the linking number. 
\par 
 Let us now apply the gauge transformation (\ref{gt-vertex})
with $\epsilon=1$ and $\kappa=1/2$ to the weights 
given by eqs. (\ref{abcRmatrix}) \cite{AW} . Then,  
from the transformed Boltzmann weights 
we have 
\be
\lim_{z \rightarrow \infty} {\tilde X}(z)/\sinh(z + 2 \eta) = \left(
\begin{array}{cccc} 
1 &   0 & 0 & 0 \\
0 &  0 & \exp(- 2 \eta) & 0 \\
0 &   \exp(- 2 \eta) & 1 - \exp(- 4 \eta) & 0 \\
0 &   0 &  0  &  1
\ear
\right) 
\label{bm2} 
\ee
This matrix representation of the braid group 
leads to the Jones polynomial with 
$q=\exp(2 \eta)$ \cite{AW}.

\par We remark that 
 based on  the representations of the braid group which are 
derived from the Boltzmann weights of exactly solvable models, 
 we can construct various invariants of knots and links 
 (see, for reviews, Refs. \cite{WDA,Kauffman,FYWu}).

\subsection{Quantum groups (Hopf algebras)}

From the quantum groups (the Hopf algebras) we can systematically construct 
representations of the braid group  
such as derived  from the six-vertex model.   
Almost all the solutions of the Yang-Baxter equations 
can be constructed in some framework of quantum groups.   
Furthermore, the connection of solvable models to the quantum groups 
is useful for investigating nontrivial properties of integrable models.

\par 
Let us introduce the quantum group $U_q(sl_2)$ , which is a $q$-analog of 
the universal enveloping algebra of $sl_2$. 
Generators $X^{\pm}, H$ satisfy 
\be 
K X^{\pm} K^{-1} = q^{\pm 2} X^{\pm } \, , \quad 
 [X^{+}, X^{-}] = {\frac {K - K^{-1}} {q-q^{-1}}}  
\ee
We may express $K$ as $K=q^H$. Taking the limit of $q$ to unity, 
the relations are reduced into the commutation relations of $sl_2$. 

\par 
The tensor product is defined by the following 
\bea 
\Delta(K) & = & K \otimes K \non \\
\Delta(X^{+}) & = & X^{+} \otimes I + K \otimes X^{+} \non \\
\Delta(X^{-}) & = & X^{-} \otimes K^{-1} + I \otimes X^{-} 
\eea
The operation $\Delta(\cdot)$ is called the comultiplication. 
In the quantum group, the comultiplication does not commute with the 
exchange operator  $\sigma$ defined by 
$\sigma a \otimes b = b\otimes a$. However, there is 
an operator $R$ which satisfy the following 
\be
R \Delta(x) = \sigma \Delta(x) R \, , \quad {\rm for} \quad 
x \in U_q(sl_2) \label{int}
\ee
Thus, the tensor product $V_1 \otimes V_2$ can be related to 
$V_2 \otimes V_1$ through the R matrix of the quantum group. 

\par 
In the $U_q(sl_2)$, the R matrix can be constructed in the operator formalism. 
If operators $X^{\pm}$ and $H$ satisfy the defining relations of the 
$U_q(sl_2)$, then the operator $\cal R$ defined by the following 
satisfy the intertwining relation (\ref{int})
\be 
{\cal R} = q^{-H \otimes H /2} 
\exp_q \left( -(q-q^{-1}) K^{-1} X^{+} \otimes X^{-} K \right) 
\ee
where $\exp_q x$ denotes the infinite series 
\be
\exp_q x = \sum_{n=0}^{\infty} q^{-n(n-1)/2} {\frac {x^n} {[n]!} }  
\ee
The operator ${\cal R}$ is called the universal $R$ matrix. 

\par 
The  representation (\ref{bm2}) of the braid group 
corresponds to the representation of the 
universal $R$ matrix on the tensor product 
of two fundamental representations. 

\par 
We note that the universal $R$ matrix can be constructed canonically 
through  Drinfeld's quantum double construction 
(for instance, see \cite{Etingof}).  This is  similar to  
 the Sugawara construction which derives 
the energy momentum tensor 
from the current operator.

%
%

%
%

\setcounter{equation}{0} 
\setcounter{figure}{0} 
\renewcommand{\theequation}{A.\arabic{equation}}
\renewcommand{\thefigure}{A.\arabic{figure}}
\setcounter{section}{0}

\appendix
\section{Commuting transfer matrices and the Yang-Baxter equations }

We show that if given two sets of the Boltzmann weights 
of the six vertex model 
satisfy the Yang-Baxter relation, then their transfer matrices commute.  
We consider three sets of the Boltzmann weights:  
$(w_1,w_2,w_3)=(a,b,c)$, $(a^{'}, b^{'}, c^{'})$, and 
$(a^{''}, b^{''}, c^{''})$. 
Let us  denote by $\tau^{'}$ and $\tau^{''}$ the transfer matrices 
constructed from the sets of the Boltzmann weights 
 $(a^{'}, b^{'}, c^{'})$ and $(a^{''}, b^{''}, c^{''})$, respectively.  
Then, we can show that 
if the three sets of the Boltzmann weights satisfy the Yang-Baxter equations 
given in \S 2.3, then the transfer matrices 
$\tau^{'}$ and $\tau^{''}$ commute. 

\par 
Let us now explicitly discuss the commutation relation. 
We first introduce the monodromy matrix. It is an $N$ ranked tensor, 
whose $(\al, \bt)$ elements are defined as follows  
\be \left( 
 T_{\al,\bt} \right)^{a_1, \ldots, a_N}_{b_1, \ldots, b_N} 
= \sum_{c_2, \ldots, c_N} w(\beta, b_1 |  a_1, c_2) 
 w(c_2, b_2 | a_2, c_3 ) 
\cdots w(c_N, b_N | a_N, \alpha)  
\ee
The transfer matrix is given by the trace of the monodromy matrix 
\be 
\tau = {\rm tr} \left( T \right) = \sum_{\al=1,2} T_{\al,\al} 
\ee
In terms of the matrix elements, we have 
\be \left( 
 \tau \right)^{a_1, \ldots, a_N}_{b_1, \ldots, b_N}  
= \sum_{\al=1,2}  \left( T_{\al,\al} \right)
^{a_1, \ldots, a_N}_{b_1, \ldots, b_N}  
\ee

%
%
\begin{figure}[htb]
%
\begin{center}
\setlength{\unitlength}{1mm}
\begin{picture}(160,20)(-30,10)
\put(10,20){\thicklines\line(1,0){80}}
\multiput(20,20)(15,0){2}{{\thicklines\line(0,1){10}}}
\multiput(20,20)(15,0){2}{{\thicklines\line(0,-1){10}}}
\put(80,20){\thicklines\line(0,1){10}}
\put(80,20){\thicklines\line(0,-1){10}}
\put(5,20){\makebox(5,5){\LARGE $\beta$}}  
\put(90,20){\makebox(5,5){\LARGE $\alpha$}} 
\put(25,20){\makebox(5,5){$c_2$}} 
\put(70,20){\makebox(5,5){$c_{N}$}} 
\put(55,25){\makebox{$\cdots$}}
\put(55,15){\makebox{$\cdots$}}
\put(20,30){\makebox(5,5){$a_1$}}  
\put(20,5){\makebox(5,5){$b_1$}} 
\put(35,30){\makebox(5,5){$a_2$}}  
\put(35,5){\makebox(5,5){$b_2$}} 
\put(80,30){\makebox(5,5){$a_N$}}  
\put(80,5){\makebox(5,5){$b_N$}} 
\end{picture} 
\end{center}
\caption{The matrix element $({\alpha}, {\beta})$ of the monodromy matrix¡§
$\left( T_{\al,\bt} \right)^{a_1, \ldots, a_N}_{b_1,\ldots, b_N}$} 
\end{figure}

\par 
Let us denote by $T^{'}$ and $T^{''}$ the monodromy matrices 
for the sets of the Boltzmann weights 
$(a^{'}, b^{'}, c^{'})$ and $(a^{''}, b^{''}, c^{''})$, respectively.  
We consider the product of the matrix elements of the two monodromy matrices:
 $T^{'}_{\al,\bt} T^{''}_{\gm,\dt}$.  
The entry of $(a_1, \ldots , a_N)$ and $(b_1, \ldots, b_N)$ 
of the product $T^{'}_{\al,\bt} T^{''}_{\gm,\dt}$   is given by 
\bea
& & \left( T^{'}_{\gm_1,\gm_{N+1} } T^{''}_{\dt_1,\dt_{N+1}} \right)
^{a_1, \ldots, a_N}_{b_1, \ldots, b_N} 
 = \sum_{e_1, \ldots, e_N} \left(
 T^{'}_{\gm_1, \gm_{N+1} }
\right)^{a_1, \ldots, a_N}_{e_1, \ldots, e_N} \left(
 T^{''}_{\dt_1, \dt_{N+1} } \right)
 ^{e_1, \ldots, e_N}_{b_1, \ldots, b_N} 
\non \\
&=& \sum_{e_1, \ldots, e_N} \sum_{c_2, \ldots, c_N} 
 w^{'}(\gm_1, e_1 | a_1, c_2) w^{'}(c_2, e_2 | a_2, c_3) 
    \cdots w^{'}(c_N, e_N | a_N, \gm_{N+1}) 
\non \\
& & \qquad \times \sum_{d_2, \ldots, d_N}
w^{''}(\dt_1, b_1 | e_1, d_2) w^{''}(d_2, b_2 | e_2, d_3) 
\cdots w^{''}(d_N, b_N | e_N, \dt_{N+1}) 
\non \\
&=&  \sum_{c_2, \ldots, c_N} 
\sum_{d_2, \ldots, d_N} \sum_{e_1, \ldots, e_N}
\left( w^{'}(\gm_1, e_1 | a_1, c_2) w^{''}(\dt_1, b_1 | e_1, d_2) \right) 
\cdot 
\non \\
& & \left( w^{'}(c_2, e_2 | a_2, c_3) w^{''}(d_2, b_2 | e_2, d_3) \right)
  \cdots \left( w^{'}(c_N, e_N | a_N, \gm_{N+1}) 
w^{''}(d_N, b_N | e_N, \dt_{N+1}) \right)
\non \\
&=&  \sum_{c_2, \ldots, c_N} 
\sum_{d_2, \ldots, d_N} 
S(a_1, b_1)^{ \gm_1,c_2}_{\dt_1,d_2}
\cdot 
S(a_2, b_2)^{c_2,c_3}_{d_2,d_3} \cdots 
S(a_N, b_N)^{c_N, \gm_{N+1}}_{d_N, \dt_{N+1}}   
\eea
Here $S(a_j, b_j)^{c_j, c_{j+1}}_{d_j, d_{j+1}}$ has been defined by  
\be 
S(a_j, b_j)^{c_j, c_{j+1}}_{d_j, d_{j+1}} = 
 \sum_{e_j} w^{'}(c_j, e_j | a_j, c_{j+1}) w^{''}(d_j, b_j | e_j, d_{j+1}) 
\ee
We define the matrix element $M^{c_0, c_1}_{d_0,d_1}$  
as follows 
\be 
M^{c_0, c_1}_{d_0,d_1} = w(d_0, d_1 | c_0, c_1) 
\ee
Here we assume that  $M^{c_0, c_1}_{d_0,d_1}$  
denotes the matrix element 
for column $(c_0, d_0)$ and row $(c_1,d_1)$ 
of the matrix $M$. 
Multiplying the matrix $M$ to the product $T^{'}_{\al,\bt} T^{''}_{\gm,\dt}$   
and applying the Yang-Baxter relation $N$ times,  
we can derive  the following 
\be 
\sum_{c_1,d_1} M^{\gm_0, c_1}_{\dt_0,d_1} 
T^{'}_{c_1, \gm_{N+1}} T^{''}_{d_1, \dt_{N+1}} 
=
\sum_{c_{N},d_{N}} 
T^{''}_{c_1, c_{N+1}} T^{'}_{d_1, d_{N+1}} 
M^{c_{N}, \gm_{N+1}}_{d_{N}, \dt_{N+1}} 
\label{rail} 
\ee

%
%
\begin{figure}[htb]
\begin{center}
\setlength{\unitlength}{0.7mm}
\begin{picture}(180,80)(10,5)
\put(10,90){\thicklines\line(1,-1){20}}
\put(10,70){\thicklines\line(1,1){20}}
\put(30,90){\thicklines\line(1,0){50}}
\put(30,70){\thicklines\line(1,0){50}}
\put(35,65){\thicklines\line(0,1){30}}
\put(45,65){\thicklines\line(0,1){30}}
\put(75,65){\thicklines\line(0,1){30}}
\put(35,95){\makebox(5,5){$a_1$}}
\put(45,95){\makebox(5,5){$a_2$}}
\put(75,95){\makebox(5,5){$a_N$}}
\put(35,60){\makebox(5,5){$b_1$}}
\put(45,60){\makebox(5,5){$b_2$}}
\put(75,60){\makebox(5,5){$b_N$}}
\put(60,95){\makebox(5,5){$\cdots$}} 
\put(60,75){\makebox(5,5){$\cdots$}} 
\put(60,60){\makebox(5,5){$\cdots$}} 
\put(90,75){\makebox(5,5){$=$}} 
\put(100,90){\thicklines\line(1,0){10}}
\put(100,70){\thicklines\line(1,0){10}}
\put(110,90){\thicklines\line(1,-1){20}}
\put(110,70){\thicklines\line(1,1){20}}
\put(130,90){\thicklines\line(1,0){40}}
\put(130,70){\thicklines\line(1,0){40}}
\put(105,65){\thicklines\line(0,1){30}}
\put(135,65){\thicklines\line(0,1){30}}
\put(165,65){\thicklines\line(0,1){30}}
\put(105,95){\makebox(5,5){$a_1$}}
\put(135,95){\makebox(5,5){$a_2$}}
\put(165,95){\makebox(5,5){$a_N$}}
\put(105,60){\makebox(5,5){$b_1$}}
\put(135,60){\makebox(5,5){$b_2$}}
\put(165,60){\makebox(5,5){$b_N$}}
\put(150,95){\makebox(5,5){$\cdots$}} 
\put(150,75){\makebox(5,5){$\cdots$}} 
\put(150,60){\makebox(5,5){$\cdots$}} 
\put(90,45){\makebox(5,5){$\cdot$}}
\put(90,50){\makebox(5,5){$\cdot$}}
\put(90,55){\makebox(5,5){$\cdot$}}
\put(90,20){\makebox(5,5){$=$}} 
\put(100,30){\thicklines\line(1,0){50}}
\put(100,10){\thicklines\line(1,0){50}}
\put(150,30){\thicklines\line(1,-1){20}}
\put(150,10){\thicklines\line(1,1){20}}
\put(105,5){\thicklines\line(0,1){30}}
\put(115,5){\thicklines\line(0,1){30}}
\put(145,5){\thicklines\line(0,1){30}}
\put(105,35){\makebox(5,5){$a_1$}}
\put(115,35){\makebox(5,5){$a_2$}}
\put(145,35){\makebox(5,5){$a_N$}}
\put(105,0){\makebox(5,5){$b_1$}}
\put(115,0){\makebox(5,5){$b_2$}}
\put(145,0){\makebox(5,5){$b_N$}}
\put(130,35){\makebox(5,5){$\cdots$}} 
\put(130,15){\makebox(5,5){$\cdots$}} 
\put(130,0){\makebox(5,5){$\cdots$}} 
\put(5,90){\makebox(5,5){$\gamma_0$}} 
\put(25,90){\makebox(5,5){$c_1$}} 
\put(38,90){\makebox(5,5){$c_2$}} 
\put(82,90){\makebox(5,5){$\gamma_{N+1}$}} 
\put(95,90){\makebox(5,5){$\gamma_0$}} 
\put(112,90){\makebox(5,5){$c_1$}} 
\put(123,90){\makebox(5,5){$c_2$}} 
\put(170,90){\makebox(5,5){$\gamma_{N+1}$}} 
\put(5,65){\makebox(5,5){$\delta_0$}} 
\put(25,65){\makebox(5,5){$d_1$}} 
\put(38,65){\makebox(5,5){$d_2$}} 
\put(82,65){\makebox(5,5){$\delta_{N+1}$}} 
\put(95,65){\makebox(5,5){$\delta_0$}} 
\put(112,65){\makebox(5,5){$d_1$}} 
\put(123,65){\makebox(5,5){$d_2$}} 
\put(170,65){\makebox(5,5){$\delta_{N+1}$}} 
\put(95,30){\makebox(5,5){$\gamma_0$}} 
\put(108,30){\makebox(5,5){$c_1$}} 
\put(153,30){\makebox(5,5){$c_N$}} 
\put(170,30){\makebox(5,5){$\gamma_{N+1}$}} 
\put(95,5){\makebox(5,5){$\delta_0$}} 
\put(108,5){\makebox(5,5){$d_1$}} 
\put(153,5){\makebox(5,5){$d_N$}} 
\put(170,5){\makebox(5,5){$\delta_{N+1}$}} 
\put(35,90){\circle{2}}
\put(45,90){\circle{2}}
\put(75,90){\circle{2}}
\put(35,70){\circle*{2}}
\put(45,70){\circle*{2}}
\put(75,70){\circle*{2}}
\put(105,70){\circle{2}}
\put(135,90){\circle{2}}
\put(165,90){\circle{2}}
\put(105,90){\circle*{2}}
\put(135,70){\circle*{2}}
\put(165,70){\circle*{2}}
\put(105,30){\circle*{2}}
\put(115,30){\circle*{2}}
\put(145,30){\circle*{2}}
\put(105,10){\circle{2}}
\put(115,10){\circle{2}}
\put(145,10){\circle{2}}
\end{picture}
\end{center}
\caption{Pictorial proof of the commutation relation: 
$MT^{'}T^{''} = T^{''}T^{'} M$.   Open and closed circles
denote the Boltzmann weights $w^{'}$ and $w^{''}$, respectively.
 The summation over variables $c_1, \ldots, c_N$ and $d_1, \ldots, d_N$ 
 is assumed. }  
\end{figure}

\par 
Let us briefly discuss the derivation of relation 
(\ref{rail}).  It is depicted in Fig. A.2. 
In the first equality of Fig. 2.6, we have applied the Yang-Baxter relation 
formulated as follows 
\be
\sum_{c_1,d_1} M^{\gm_0,c_1}_{\dt_0, d_1} 
 S(a_1,b_1)^{c_1,c_2}_{d_1,d_2} 
= 
\sum_{c_1,d_1}  S^{'}(a_1,b_1)^{\gm_0,c_1}_{\dt_0,d_1} 
 M^{c_1,c_2}_{d_1, d_2}
\ee
Here the symbol $S^{'}(a_j, b_j)^{c_j, c_{j+1}}_{d_j, d_{j+1}}$ has been defined by 
\be 
S^{'}(a_j, b_j)^{c_j, c_{j+1}}_{d_j, d_{j+1}} 
= \sum_{e_j} w^{''}(c_j, e_j | a_j, c_{j+1}) 
w^{'}(d_j, b_j | e_j, d_{j+1}) 
\ee
We also note that the LHS of (\ref{rail}) corresponds to the sum  
\be 
 \sum_{c_1,c_2, \ldots, c_N} 
\sum_{d_1, d_2, \ldots, d_N} 
M^{\gm_0, c_1}_{\dt_0,d_1}
S(a_1, b_1)^{ c_1,c_2}_{d_1,d_2}
\cdot 
S(a_2, b_2)^{c_2,c_3}_{d_2,d_3} \cdots 
S(a_N, b_N)^{c_N, \gm_{N+1}}_{d_N, \dt_{N+1}}   
\ee

\par 
Let us consider the inverse of the matrix $M$
\be 
\left(
 M^{-1} M \right)^{c_1, c_2}_{d_1, d_2}
= \left( M M^{-1} \right)^{c_1, c_2}_{d_1, d_2}
 = \delta_{c_1,c_2} \delta_{d_1,d_2} 
\ee
Multiplying the inverse $M^{-1}$ to both hand sides of 
(\ref{rail}) we have 
\be 
M T^{'}T^{''} M^{-1} = T^{''} T^{'}
\ee
Noting $tr(M T^{'}T^{''} M^{-1} )= tr( T^{'}T^{''} )$, we obtain 
the commutation relation of the transfer matrices 
\be 
\tau^{'}\tau^{''} = \tau^{''}\tau^{'}
\ee

\end{document}